\documentclass[fleqn,usenatbib]{mnras}

\usepackage{newtxtext,newtxmath}
\usepackage[T1]{fontenc}
\usepackage{graphicx}	
\usepackage{amsmath}	
\usepackage[usenames,dvipsnames]{xcolor}
\usepackage{hyperref}
\usepackage[normalem]{ulem}
\usepackage{soul}
\usepackage[dvipsnames]{xcolor}

\usepackage{macros}

\newcommand{\sub}[1]{_{\mathrm{#1}}}
\newcommand{\msun}{M\sub{\sun}}
\newcommand{\nbody}{$N$-body }
\newcommand{\sersic}{S{\'e}rsic }
\newcommand{\Mdet}{M_{\ast}^{\rm obs}}
\newcommand{\Mbound}{M_{\ast}}

\def\equationautorefname~#1\null{Eq.~(#1)\null}
\def\figureautorefname~#1\null{Fig.~#1\null}
\newcommand{\appref}[1]{\hyperref[#1]{Appendix~\ref{#1}}}

\title[Tidal interactions and DM deficient galaxies]
      {On the Tidal Formation of Dark Matter Deficient Galaxies}

\author[G. Ogiya et al.]{
Go Ogiya$^{1,2}$\thanks{E-mail: gogiya@uwaterloo.ca (GO)},
Frank~C.~van den Bosch$^{3}$, and 
Andreas Burkert$^{4,5}$
\vspace*{8pt}
\\
$^{1}$Waterloo Centre for Astrophysics, University of Waterloo, Waterloo, ON N2L 3G1, Canada \\
$^{2}$Department of Physics and Astronomy, University of Waterloo, 200 University Avenue West, Waterloo, Ontario N2L 3G1, Canada \\
$^{3}$Department of Astronomy, Yale University, PO. Box 208101, New Haven, CT 06520-8101 \\
$^{4}$Universit\"ats-Sternwarte M\"unchen, Scheinerstra\ss e 1, D-81679 M\"unchen, Germany \\
$^{5}$Max-Planck-Institut f\"ur extraterrestrische Physik, Postfach 1312, Gie\ss enbachstra\ss e, D-85741 Garching, Germany 
}

\date{Accepted XXX. Received YYY; in original form ZZZ}

\pubyear{2021}

\begin{document}

\label{firstpage}
\pagerange{\pageref{firstpage}--\pageref{lastpage}}
\maketitle


\begin{abstract}
  Previous studies have shown that dark matter deficient galaxies (DMDG) such as NGC1052-DF2 (hereafter DF2) can result from tidal stripping. An important question, though, is whether such a stripping scenario can explain DF2’s large specific frequency of globular clusters (GCs). After all, tidal stripping and shocking preferentially remove matter from the outskirts. We examine this using idealized, high-resolution simulations of a regular dark matter dominated galaxy that is accreted onto a massive halo. As long as the initial (pre-infall) dark matter halo of the satellite is cored, which is consistent with predictions of cosmological, hydrodynamical simulations, the tidal remnant can be made to resemble DF2 in all its properties, including its GC population. The required orbit has a peri-centre at the 8.3 percentile of the distribution for subhaloes at infall, and thus is not particularly extreme. On this orbit the satellite loses 98.5 (30) percent of its original dark matter (stellar) mass, and thus evolves into a DMDG. The fraction of GCs that is stripped off depends on the initial radial distribution. If, at infall, the median projected radius of the GC population is roughly two times that of the stars, consistent with observations of isolated galaxies, only $\sim 20$ percent of the GCs are stripped off. This is less than for the stars, which is due to dynamical friction counteracting the tidal stirring. We predict that, if indeed DF2 was crafted by strong tides, its stellar outskirts should have a very shallow metallicity gradient.
\end{abstract}



\section{Introduction}
\label{sec:intro}

Recently, \cite{vanDokkum2018_df2} reported that the ultra diffuse galaxy NGC~1052-DF2 (hereafter DF2), believed to be a member of the NGC~1052 group, is dark matter (DM) deficient. In particular, its inferred DM mass ($M_\rmh < 1.5 \times 10^8\Msun$ at 90\% CL) is smaller, by a factor of a few hundred, than what is expected ($M_\rmh \sim 6 \times 10^{10} \Msun$) based on its stellar mass ($M_\ast \sim 2 \times 10^8 \Msun$) and empirical models for galaxy formation and evolution  \citep[e.g.,][]{Moster2018, Behroozi2019}. This inference was originally based on the extremely small line-of-sight velocity dispersion ($8.5^{+2.3}_{-3.1} \kms$) among the 10 globular clusters (GCs) that orbit DF2. Although inferring a mass from this measurement has its shortcomings \citep[][]{Martin2018, Hayashi2018, Laporte2019, Lewis2020, Montes2021}, the inference that DF2 is DM deficient has subsequently been verified by more sophisticated Jeans modelling \citep{Wasserman2018} and by the measurements and modelling of the kinematics of stars and planetary nebulae in DF2 \citep{Danieli2019, Emsellem2019}. Yet, the DM halo mass, $M_\rmh$, that one infers from the observed {\it number} of globular clusters, $N\sub{GC}$, based on the empirical $M_\rmh$-$N\sub{GC}$ relation of \citet{Burkert2020} is $M_\rmh \sim 5 \times 10^{10} \Msun$, which is perfectly consistent with standard expectations based on the empirical galaxy formation models with stellar masses as observed for DF2. The inferred luminosity function of these GCs, though, is abnormal, in that it appears offset towards the bright end compared to other galaxies \citep{vanDokkum2018_gc, Shen2021_gc}. In fact, DF2 seems to have several GCs that are comparable in size and luminosity to $\omega$-Cen, the biggest and brightest GC in the Milky Way. To add to the puzzle, in 2019 a second galaxy in the same field, NGC~1052-DF4 (hereafter DF4), was found to have properties that are remarkably similar to DF2, including an inferred DM deficiency \citep[][]{vanDokkum2019}.

The claim for DM deficiency depends critically on the distances to DF2 and DF4. In two studies, \cite{Trujillo2019} and \cite{Monelli2019} claim that the distances to DF2 and DF4 are $\sim 13-14\Mpc$, significantly less than the distance of $\sim 20\Mpc$ advocated by \cite{vanDokkum2018_df2} and \cite{vanDokkum2019}. If true, this would make DF2 and DF4 far more ordinary, similar to nearby dwarf galaxies, with no notable DM deficiency, and with normal GC luminosity functions. However, recent analyses of the tip of the red giant branch, based on deep Hubble Space Telescope photometry by \cite{Danieli2020} and \cite{Shen2021_distance}, seem to confirm the larger distances, and thus the deficiency of DM in both DF2 and DF4.

The existence of dark matter deficient galaxies (hereafter DMDGs) has long been anticipated. As first suggested by \cite{Zwicky.56} and \cite{Schweizer.78}, dwarf galaxies can form out of the tidal debris (tidal arms) that form when two gas-rich galaxies merge together. These ideas received little attention, though, until \cite{Mirabel.etal.92} reported the discovery of a tidal dwarf galaxy (TDG) forming in the outskirts of the tidal arms of the Antennae galaxy. Since then, the existence of TDGs has become well established observationally \citep[e.g.,][and references therein]{Braine.etal.00, Weilbacher.etal.00, Duc.etal.00}. Starting with \citet{Barnes.Hernquist.92}, numerical simulations of gas-rich galaxy mergers have shown that TDGs are expected to be DM deficient \citep[see also e.g.,][]{Elmegreen.etal.93, Bournaud.Duc.06, Wetzstein.etal.07, Ploeckinger.etal.18}. Observations of the HI rotation curves of some TDGs (still embedded in their tidal debris) has indeed confirmed that their dynamical mass is comparable to their baryonic mass \citep[e.g.,][]{Bournaud.etal.07, Lelli.etal.15}. Although it is unclear how common TDGs are, \cite{Okazaki.Taniguchi.00} have argued that if only a few dwarf galaxies are formed in each galaxy collision, the expected number density of TDGs could be comparable to that of dwarf ellipticals. On the other hand, \cite{Kaviraj2012} used a large sample of galaxy mergers and inferred that only $\sim 6$ percent of dwarf galaxies in local clusters may be of tidal origin.

Clearly, then, it is tempting to interpret DF2 and DF4 as examples of older TDGs that are no longer embedded in the tidal debris out of which they formed. A key signature of such systems, though, is that, since they form out of strongly metal-enriched material, they should be outliers of the mass-metallicity relation. Such outliers are known to exist and typically identified as candidate TDGs \citep[e.g.,][]{Hunter.etal.00, Sweet.etal.14, Duc.etal.14}. However, spectra with MUSE on the VLT indicate that both the stellar body and the GCs of DF2 have low metallicity, in agreement with the mass-metallicity relation \citep[][]{Fensch.etal.19}. This makes it unlikely that DF2 is a TDG, motivating an exploration of alternative formation scenarios.

One such scenario was proposed by \cite{Silk2019}, who suggested that DF2 can be the result of a mini Bullet-cluster event involving a high-speed collision between two gas-rich dwarf galaxies. While this compresses the baryonic gas, triggering the formation of over-pressured dense clouds that are inducive to GC formation, the DM components pass through each other and separate from the baryonic material, thus giving rise to a DMDG. While this scenario plays out in numerical simulations \citep[see e.g.,][]{Shin.etal.20, Lee.etal.21}, the conditions that give rise to this formation channel should be rare, making it unlikely that one galaxy group (NGC-1052) hosts two such systems (DF2 plus DF4).

Another alternative formation scenario for DMDGs was proposed by \cite{Ogiya2018} and \cite{Nusser.20}, who argue that DF2 and DF4 might be extreme examples of tidal remnants. In this scenario, the DMDGs start out as normal galaxies whose DM halos are largely stripped off by tidal forces after they are accreted into a bigger halo (in the case of DF2 and DF4 this is the host halo of NGC~1052). Using high-resolution, idealized simulations, \cite{Ogiya2018} showed that, for sufficiently radial orbits, it is indeed possible to preferentially remove DM from the satellite and create a DMDG that resembles DF2 \citep[see also][]{Yang2020, Maccio2021}. A key requirement, though, is that the progenitor has a cored DM profile. If DM is self-interacting, such a core is a natural outcome \citep[e.g.,][]{Spergel.Steinhardt.00, Elbert.15}. In the case of $\Lambda$ Cold Dark Matter ($\Lambda$CDM), cores can be created due to processes that deplete the central DM density, such as stochastic supernova feedback \citep[e.g.,][]{Pontzen2012, DiCintio2014, Ogiya&Mori2014}. In fact, using {\tt NewHorizon}, a high-resolution cosmological simulation, \citet{Jackson2021} have demonstrated that the creation of DMDGs via this mechanism is a natural by-product of galaxy evolution in $\Lambda$CDM, and that the existence of objects such as DF2 and DF4 is therefore perfectly consistent with the standard paradigm. In addition, there is observational evidence to support the notion that DF2 and/or DF4 have indeed been influenced by strong tides \citep[][]{Montes2020, Keim2021}. 

A potential key to deciphering the formation of DF2 is its extraordinary GC population, in particular its spatial distribution and abnormal luminosity function. One potential explanation for the overly massive GCs might be that they formed due to GC-GC mergers. After all, DF2 appears conducive to such mergers since the velocity dispersion among its GCs is comparable to their internal velocity dispersion. However, using high-resolution numerical simulations, \cite{DuttaChowdhury2020} showed that the merger rate is far too low to explain the abnormal luminosity function. An alternative explanation was put forward by \cite{Trujillo-Gomez2021}, who argue that the formation of DF2 must have involved a phase of large gas compression, which might thus favour the mini Bullet-cluster formation picture of \cite{Silk2019}. 

The observed spatial distributions of GCs in DF2 and DF4, with projected half-number radii of $\sim 3\kpc$, also requires explanation. Naive predictions based on Chandrasekhar's dynamical friction formula \citep{Chandrasekhar1943} suggest that several of the observed GCs should inspiral in a few Gyr \citep[][]{Nusser2018, Leigh2020}. Unless they formed recently, which is ruled out by the fact that their stellar populations are old \citep[$8.9 \pm 1.8 \Gyr$][]{Fensch.etal.19}, this begs the question why DF2 and DF4 do not contain a nuclear star cluster. A potential solution is offered by \cite{DuttaChowdhury2019}, who used a suite of \nbody simulations including multiple GCs and found that scattering among the GCs acts as a dynamical buoyancy force that counter-balances dynamical friction. In addition, if DF2 and DF4 indeed lack DM, their cored distributions of stars imply that the orbital decay due to dynamical friction will cease close to the core radius due to an effect called `core-stalling' \citep[][]{Read.etal.06, Inoue.11, Banik2021_df}. Yet, despite these buoyancy-like processes, \cite{DuttaChowdhury2019} find that the population of GCs in DF2 should decrease its half-number radius by a factor $\sim 2$ in less than $10\Gyr$. And that is without any DM; if DF2 has a non-negligible DM halo, the inspiral rate can be substantially faster, especially if that halo is cuspy.

One of the goals of this paper is to investigate the evolution of the GC populations in DF2 and DF4 in the extreme-tide picture. In particular, using idealised, high-resolution simulations, we investigate, starting from a normal DM dominated dwarf galaxy, to what extent the tidal forces strip GCs and impact their spatial distribution due to tidal stirring, a term that we use to describe the impact of violent re-virialization in response to tidal stripping and impulsive heating. We also use the simulations to examine in detail the evolution of the size and velocity dispersion of the stellar remnant. The overarching goal is to examine whether tidal effects can transform a normal dwarf galaxy into a DMDG akin to DF2 and DF4.

This paper is organised as follows. Section~\ref{sec:sim_models} describes the $N$-body simulations and the initial conditions. Section~\ref{sec:formation_dm_deficient_galaxies} presents the results and demonstrates that tidal stripping is indeed a viable mechanism for creating DMDGs with properties similar to DF2 and DF4. In Section~\ref{sec:internal_structure} we make some predictions for the internal structure of DMDGs, and we summarise our findings in Section~\ref{sec:summary}. Throughout, where needed, we adopt cosmological parameters taken from \cite{Planck2016}. 

\section{Simulation Models}
\label{sec:sim_models}

We simulate the tidal evolution of a satellite galaxy, which is to represent a potential progenitor of DF2, as it orbits a larger host system, similar to the DM halo of the NGC~1052 group. The simulations cover the period from shortly before infall, which we take to occur at a redshift $z\sub{acc}$, to the present. The satellite galaxy is modelled as an \nbody system containing DM, stars and GCs, while a time-varying analytical potential is employed to model the host galaxy. 

Throughout, we adopt $z\sub{acc} = 1.5$, based on the assumption that a large fraction of the stellar mass of DF2 was formed prior to infall (i.e., we assume that the galaxy quenched shortly before or after accretion into the NGC~1052 system). Observations indicate that the age of the stars and GCs in DF2 is $\sim 9$\,Gyr \citep[][]{vanDokkum2018_gc, Fensch.etal.19}, which is comparable to the lookback time to $z\sub{acc}=1.5$. While tides can induce (bursty) star formation in gas-rich progenitors \citep[e.g.][]{Jackson2021_lsb}, such  enhancements typically only cause a modest increase in total stellar mass \citep{Pearson2019}. Hence, our simulation results would not change dramatically if the DF2 progenitor was still gas rich at the moment of infall.

Note that this improves on the \nbody simulations of \cite{Ogiya2018}, where it was assumed that $z\sub{acc}=0$, and the tidal evolution of the satellite was followed for a period of 10\,Gyr. This is important since orbital times at higher redshifts are shorter. Hence, the approach taken here results in more peri-centric passages, and therefore a stronger tidal impact \citep[i.e., see][]{Ogiya2021}. In addition, in order to assess the impact of tides on the GC population, we also run simulations in which the galaxy model evolves in isolation. In what follows, we refer to the simulations with and without external tidal field as {\tt Tide} and {\tt NoTide}, respectively. 


\subsection{\nbody galaxy model}
\label{ssec:sat_gal}

The initial \nbody system that represents the progenitor of DF2 consists of three components: a stellar spheroid, a spherical DM halo, and a population of GCs. As detailed below, its structural properties are taken to be representative of that of a `typical' galaxy, which means that it obeys the empirical stellar mass-to-halo mass relation, has a size characteristic of the observed size-mass relation of galaxies at the redshift of accretion, and has a median projected radius of its GC population in agreement with observations of (isolated) galaxies. 


\subsubsection{Stellar spheroid}
\label{sssec:stellar_spheroid_satellite}

We model the stellar body as a sphere with a projected surface brightness given by a \sersic profile \citep[][]{Sersic1963}, i.e.,
\begin{equation}
    \Sigma(R) \propto \exp{\biggl[-b_n \biggl(\frac{R}{R\sub{e}}\biggr)^{1/n} \biggr]},
        \label{eq:sersic}
\end{equation}
where $R$ is the projected distance from the centre of the galaxy, $R\sub{e}$ is the effective radius containing half of the total luminosity, $n$ is the \sersic index, and
\begin{equation}
    b_n = 2n - \frac{1}{3} + \frac{4}{405n} + \frac{46}{25515n^2}
        \label{eq:bn}
\end{equation}
\citep{Ciotti1999}. The corresponding 3D radial density profile is given by
\begin{equation}
    \rho\sub{*}(r) = \rho\sub{0} \biggl(\frac{r}{R\sub{e}}\biggr)^{-p_n} \exp{\biggl[-b_n \biggl(\frac{r}{R\sub{e}}\biggr)^{1/n} \biggr]}
        \label{eq:deprojected_sersic}
\end{equation}
\citep[e.g.][]{Mellier1987,Prugniel1997}, where $r$ is the three-dimensional distance from the centre of the galaxy, $\rho\sub{0}$ is a characteristic density, and
\begin{equation}
    p_n = 1 - \frac{0.6097}{n} + \frac{0.05463}{n^2}.
        \label{eq:pn}
\end{equation}
\citep[][]{LimaNeto1999}. Throughout, we adopt a total stellar mass of $M\sub{*} = 2 \times 10^8 \msun$, and a \sersic index of $n=1$, which are comparable to the present-day values of DF2. However, for the effective radius we adopt $R\sub{e} = 1.25 \kpc$, which is the typical size of a galaxy of this stellar mass at $z \sim 1.5$ \citep[e.g.,][]{vanderWel2014}.


\subsubsection{DM halo}
\label{sssec:dm_halo_satellite}

For the DM halo we consider two different spherical profiles: a standard NFW profile \citep[][]{Navarro1997}
\begin{equation}
    \rho\sub{NFW}(r) = \rho\sub{s} (r/r\sub{s})^{-1} (1+r/r\sub{s})^{-2},
        \label{eq:nfw_density_profile}
\end{equation}
where $\rho\sub{s}$ and $r\sub{s}$ are the scale density and scale radius, respectively, and a halo with a central density core. The latter is motivated by the observed rotation curves of dwarf galaxies \citep[e.g.][]{Flores1994, Burkert1995, Swaters2003, Walker2011, Burkert2015, Oh2015} and by hydrodynamical simulations that indicate that various baryonic processes can transform the NFW cusp into a central core \citep[e.g.,][]{Mashchenko2006, DiCintio2014, Onorbe2015}.

Following \cite{Read2016} we model the enclosed mass profile of the haloes as
\begin{equation}
    M\sub{DM}(r) = f^m(r) M\sub{NFW}(r)\,.
        \label{eq:transformed_mass_profile}
\end{equation}
Here $M\sub{NFW}(r)$ is the mass profile of a NFW halo, while
\begin{equation}
    f(r) = \tanh{(r/r\sub{c})},
        \label{eq:transformation_f}
\end{equation}
is a transformation function that describes the potential impact of baryons on the final DM profile. Here $r\sub{c}$ is a core radius, and the parameter $0 \leq m \leq 1$ controls the significance of the transformation process. In this paper we consider two cases: $m=0$ for which the halo is an unmodified NFW halo with a central $r^{-1}$ density cusp, and $m=1$ for which the cusp is transformed to a constant density core.

Throughout we assume that the mass of the halo (prior to any transformation), defined as the mass that encloses a radius $r_{200}$ inside of which the average density is 200 times the critical density, is given by $M_{200} = 6 \times 10^{10} \msun$, which is consistent with the expected halo mass for a galaxy with  $M\sub{*}=2 \times 10^8 \msun$ at $z=1.5$ \citep[see e.g.,][]{Behroozi2019}. For the concentration parameter we adopt $c_{200} = r_{200}/r\sub{s} = 6.6$, based on the concentration-mass-redshift relation of \cite{Ludlow2016}. Finally, for the transformed model ($m=1$), we adopt a core radius of $r\sub{c} = 1.75 \, R\sub{e} = 2.19 \kpc$, as suggested by \cite{Read2016}.
\begin{figure}
    \begin{center}
        \includegraphics[width=0.4\textwidth]{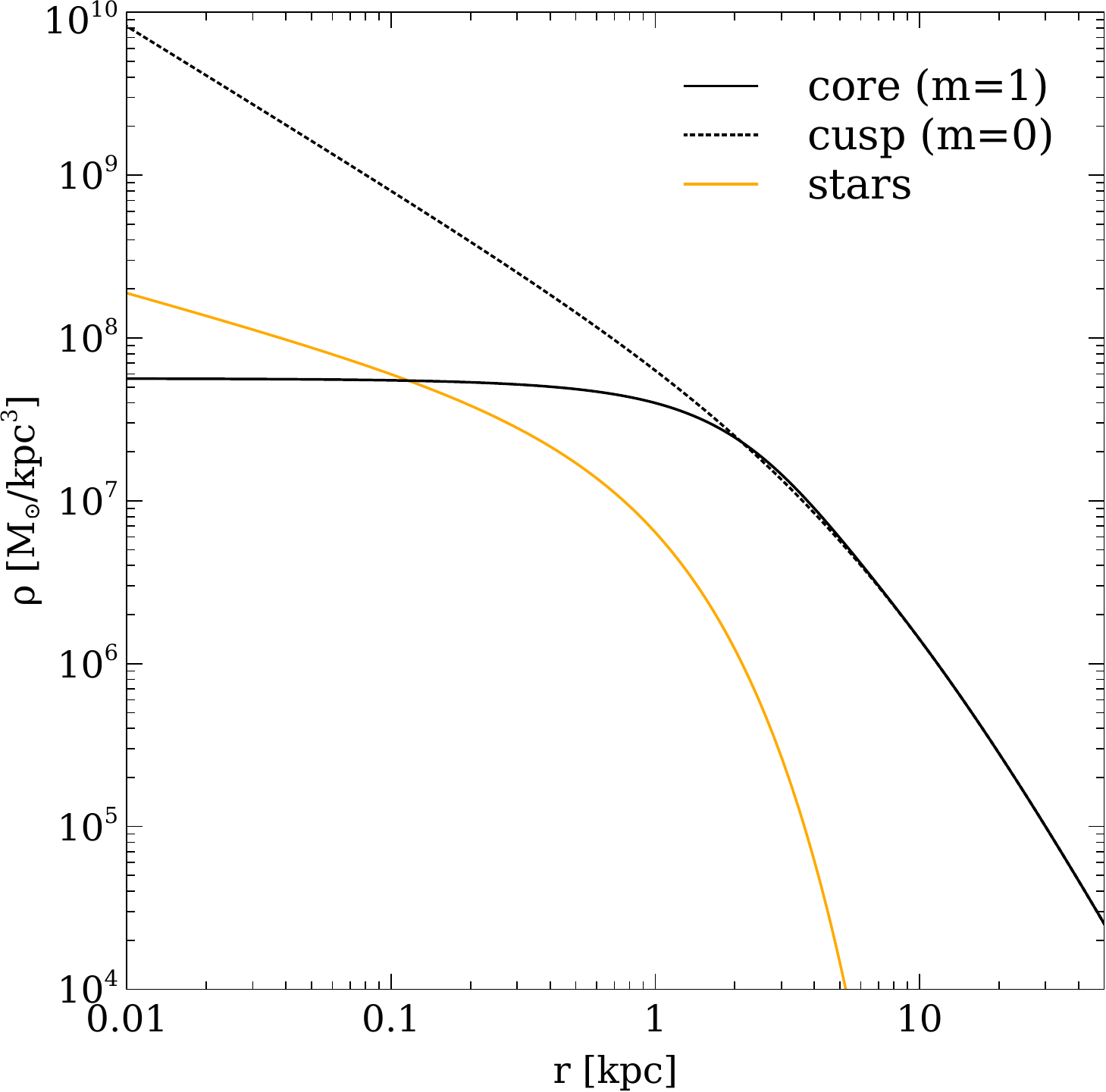}
    \end{center}
    \caption{
        Initial density profile of the \nbody satellite galaxy. Black and orange lines correspond to the DM halo and stellar spheroid, respectively. Note that we consider two different DM halos: a standard, cuspy NFW profile with $m=0$ (black dotted line) and a cored profile with $m=1$ (black solid line).
    \label{fig:density_profile}}
\end{figure}

\autoref{fig:density_profile} plots the initial density profiles of the stellar spheroid (orange) and the two DM haloes. Note that, in the case of the cored DM halo ($m=1$, solid black), the stellar density exceeds that of the DM for $r \la 0.1\kpc$. Also, the cored DM density profile is consistent with the empirical fact that the DM haloes of dwarf galaxies all seem to have a central surface density, defined as the product of the DM central density and the core radius, of $\sim 100 \Msun/\pc^2$ \citep[][and references therein]{Kormendy2016, DiPaolo2019, Burkert2020_fdm}. As shown in \cite{Ogiya2014}, this central surface density is a typical outcome of a cusp-to-core transformation process due to potential oscillations driven by star formation feedback. In the case of the cuspy NFW profile ($m=0$, dotted black), the system is DM dominated at all radii.

We draw the initial position vectors of star and DM particles using their respective density profiles and the standard random rejection sampling technique. The corresponding initial velocity vectors are stochastically sampled using the isotropic distribution functions, $f=f(E)$,  numerically computed using the Eddington formula \citep{Eddington1916} while accounting for the combined gravitational potential of the stellar spheroid and the DM halo.

\subsubsection{GCs}
\label{sssec:gc_satellite}

We select ten star particles and model them as GC particles in the simulations. The selection procedure is as follows. First, we pick a radius $r\sub{GC}$, and select the 10 star particles whose distance $r$ with respect to the centre of mass are closest to $r\sub{GC}$. We then set the masses of those particles to be $10^6 \Msun$, in rough agreement with the average GC mass in DF2 \citep[][]{vanDokkum2018_gc}, while their position and velocity vectors are unchanged from those drawn by the random rejection sampling technique.

While it may seem strange to initialize all GCs at virtually identical galacto-centric distances, this approach has several advantages. First of all, we only need a single scalar ($r\sub{GC}$) to quantify the initial conditions for the GCs. Secondly, by construction, and in the absence of dynamical friction, the GCs are in equilibrium with the stars and DM. In addition, because of their dispersion in velocities, the GC particles rapidly spread out over several kpc (i.e., they take on a more realistic radial distribution) in a timescale of $\lta 100\Myr$, which is much shorter than the simulation time or the orbital time within the host halo. In fact, after $\sim 100\Myr$ their projected number density distribution along any random orientation angle is well fit by a \sersic profile with $n=0.5$ and an effective radius $R\sub{e,GC} = 2.0 \kpc$. For comparison, the observed number density distributions of GCs are typically well fit by \sersic profiles with $n \sim 0.5 - 4$ \citep[e.g.][]{Kartha2014, Hargis2014, Kartha2016, Voggel2016}. Hence, the GC distribution in our simulations is not unrealistic, despite the somewhat artificial initialization.

In typical galaxies, the characteristic radius of the GC population, $R\sub{e,GC}$ is proportional to the effective radius of the host galaxy, albeit with an appreciable amount of scatter \citep[e.g.,][]{Forbes.17, Hudson2018}. Given that the effective radii of galaxies are proportional to their halo virial radii \citep[e.g.,][]{Kravtsov.13}, we thus expect that $R\sub{e,GC} \propto M_{200}^{1/3}$. Indeed, \citet{Forbes.17} shows that 
\begin{equation}
\log\left[\frac{R\sub{e,GC}}{\kpc}\right] = \frac{1}{3} \log\left[\frac{M_{200}}{10^{12} \Msun}\right] + 0.9
\end{equation}
provides a good match to observational data. Using that our model satellite at infall has $M_{200} = 6 \times 10^{10} \Msun$, we infer that $R\sub{e,GC} \sim 3.1 \kpc$. This is somewhat larger than the $R\sub{e,GC} = 2.0\kpc$ of our initial distribution, but the scaling relation between $R\sub{e,GC}$ and $M_{200}$ has an appreciable amount of scatter, and several systems are know that are significantly more massive than DF2 but with comparable $R\sub{e,GC}$ \citep[see][]{Forbes.17}. We briefly discuss this issue in more detail in Appendix~\ref{app:gc_orb}.

\subsection{Host galaxy}
\label{ssec:host_gal}

We model the host halo, roughly representing the DM halo of the NGC~1052 group, using an analytical, time-varying NFW potential. The NFW halo (assumed to be spherical) is fully specified by two parameters; the virial mass, $M\sub{200,host}$, and the concentration parameter $c\sub{host}$.  Each timestep in our {\tt Tide} simulations we update these structural parameters using the model for the mass assembly history of DM haloes by \cite{Correa2015} and the concentration-mass-redshift relation by \cite{Ludlow2016}. Briefly, the host mass grows roughly as $M\sub{200,host} \propto z^{-0.7}$ until $z \sim 0.7$, at which point $M\sub{200,host} \sim 7 \times 10^{12} \Msun$, followed by a reduced growth rate ($M\sub{200,host} \propto z^{-0.1}$) until it reaches $M\sub{200,host} = 1.1 \times 10^{13} \msun$ at the present time.  At $z=1.5$, when we start our simulation, the host halo mass has grown to be $\sim 3.5 \times 10^{12} \Msun$, about one-third of its final $z=0$ mass. Note that the latter is motivated by the inferred stellar mass ($\sim 10^{11}\Msun$) of NGC~1052 \citep[][]{Forbes2017} and the empirical stellar-to-halo mass relation of \cite{Moster2013}. Finally, the  halo concentration increases almost linearly with time from 4.8 at $z=1.5$ to 6.8 at $z=0$.  

\subsection{Satellite Orbit}
\label{ssec:desc_gal_interaction}

Since the host halo is assumed to be spherical, the orbit of the satellite galaxy is planar and specified by two parameters: the orbital energy $E$, and orbital angular momentum, $L$. We quantify their values with the use of two dimensionless numbers, 
\begin{equation}
    x\sub{c} \equiv r\sub{c}(E)/r\sub{200,host}(z\sub{acc})\,,
        \label{eq:xc}
\end{equation}
and 
\begin{equation}
    \eta \equiv L/L\sub{c}(E)\,.
        \label{eq:eta}
\end{equation}
Here $r\sub{c}(E)$ and $L\sub{c}(E)$ are the radius and angular momentum of a circular orbit of energy $E$, and $r\sub{200, host}(z\sub{acc})$ is the virial radius of the host halo at the time of accretion. Smaller values of $x\sub{c}$ and $\eta$ imply an orbit that is more tightly bound and more radial (i.e., $\eta = 0$ and $1$ correspond to purely radial and circular orbits, respectively).

Cosmological \nbody simulations have shown that the distributions of $x_\rmc$ and $\eta$ of subhaloes at infall are broad, with peaks around $1.2$ and $0.6$, respectively \citep[e.g.][]{Tormen1997,Zentner2005,Khochfar2006,Wetzel2011,Jiang2015,vandenBosch2018a,Li2020}. In our {\tt Tide} simulations we want to pick orbital parameters for which the tidal stripping is sufficiently strong such that the tidal remnant roughly resembles a DMDG akin to DF2 and DF4. After some trial and error (see Appendix~\ref{app:orb_params}), we settled for  $x\sub{c} = 1.0$ and $\eta = 0.3$. Although our simulations are thus tuned to produce a system that resembles DF2 (in a sense that is described below), we emphasize that these orbital parameters are by no means extreme. In particular, sampling a large number of orbits from the PDF $P(x_\rmc,\eta)$ of orbital parameters of subhaloes at infall of \cite{Jiang2015}, and analytically computing their peri-centres, we find that the peri-centre of our orbit corresponds to the 8.3 percentile of the distribution.

At the beginning of the simulations, corresponding to $z=1.5$, the centre of the $N$-body satellite system is positioned at the apocentre of its orbit with respect to the host halo with zero radial velocity. The apocentric radius and corresponding tangential bulk velocity are computed using the orbit parameters and the temporal (at $z=1.5$) properties of the host halo. Throughout the simulation the centre of the host potential is fixed at the origin. The satellite crosses the virial radius of the host system for the first time at $z \approx 1.2$. 

\subsection{Numerical parameters}
\label{ssec:num_param}

All simulations, {\tt Tide} and {\tt NoTide}, are run using an \nbody code that uses an oct-tree algorithm \citep{Barnes1986} to compute the gravitational force on each particle, accelerated with Graphics Processing Units \citep{Ogiya2013}. The stellar spheroid and DM halo of the \nbody system are modelled with 49,990 and 15,000,000 particles, respectively, resulting in a mass resolution of $\sim 4,000 \Msun$. Unless stated otherwise, we adopt a Plummer \citep{Plummer1911} force softening with softening parameter $\epsilon=14\pc$ and the cell opening criteria of \cite{Springel2005_gadget2} with the parameter controlling the force accuracy $\alpha=0.01$. Timesteps for our second-order Leapfrog scheme are updated following the prescription of \cite{Power2003} and are equal for all particles. Based on a number of tests with different numbers of particles (i.e., mass resolution) and different force softening we conclude that our simulations are numerically converged. In addition, all {\tt Tide} simulations satisfy the numerical criteria of \cite{vandenBosch2018b} assessing the reliability of simulated substructures. When analyzing the simulations, the centre and bulk velocity of the \nbody system are tracked using the iterative method outlined in \cite{vandenBosch2018a}. 

\section{Results}
\label{sec:formation_dm_deficient_galaxies}

In this section, we study the role of tidal interactions in forming DMDGs. The results from {\tt Tide} simulations, in which the \nbody galaxy model orbits the potential of the host halo, are compared to those from {\tt NoTide} simulations, in which the \nbody galaxy model is evolved in isolation. Throughout, the GCs of the satellite model are initialised using $r\sub{GC} = 2.5\kpc$, and all {\tt Tide} simulations adopt the same orbital parameters ($x\sub{c}=1.0$ and $\eta=0.3$). The only parameter that we vary is the shape parameter $m$, which characterises the central density structure of the DM halo of the satellite galaxy ($m=0$ for a cuspy profile, and $m=1$ for a cored profile, see \autoref{fig:density_profile}). All other structural parameters (e.g., initial mass, size, halo concentration, and \sersic index) of both host and satellite are kept fixed to the values discussed in \autoref{sec:sim_models}. 

\subsection{Orbital evolution and mass loss}
\label{ssec:orbit_mass}

\begin{figure}
    \begin{center}
        \includegraphics[width=0.4\textwidth]{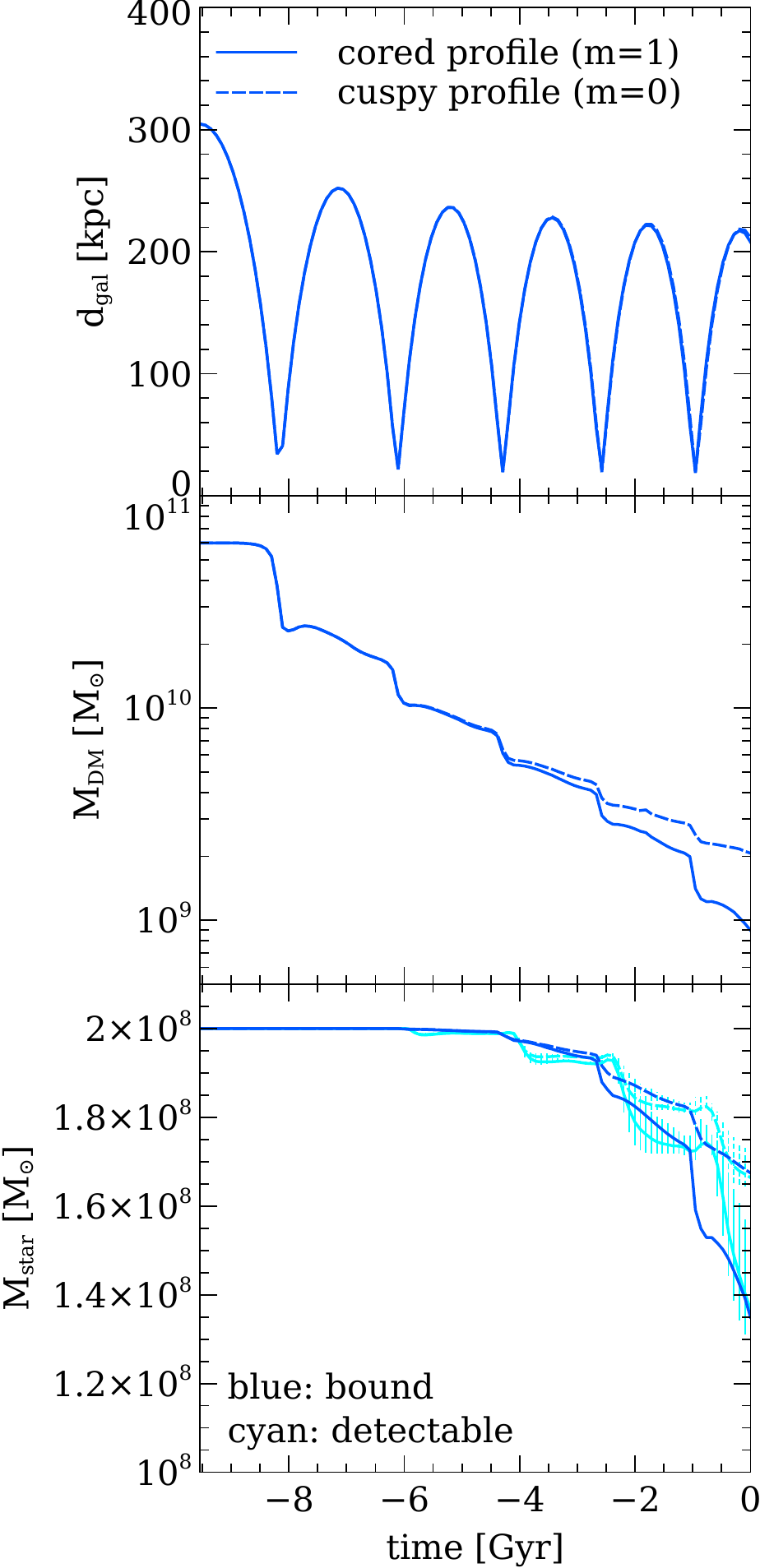}
    \end{center}
    \caption{Orbital and mass evolution of the \nbody satellite galaxy in {\tt Tide} simulations. Solid and dashed lines show the results from simulations with a cored DM halo ($m=1$) and a cuspy NFW halo ($m=0$), respectively. {\it Top panel:} distance from the centre of the host halo to the centre-of-mass of the satellite galaxy. Note that the two satellites with different halos have virtually identical orbits within their host. {\it Middle panel:} DM mass that is bound to the satellite as function of time. Note how at late times the tidal forces of the host halo strip more matter from the cored system than from its cuspy equivalent. {\it Bottom panel:} stellar mass that is bound to the satellite ($\Mbound$, blue curves) and the detectable stellar mass ($\Mdet$, cyan curves). For the latter, the errorbars indicate the 15-85 percentile range for the values obtained using 100 random orientation angles. See text for discussion.
    \label{fig:orbit_mass}}
\end{figure}

The top panel of \autoref{fig:orbit_mass} shows the evolution of the host halo-centric distance, $d\sub{gal}$, of the centre-of-mass of the satellite galaxy in our {\tt Tide} simulation. Solid and dashed curves correspond to $m=1$ and $m=0$, respectively, though their $d\sub{gal}(t)$ curves are indistinguishable. Note that the apocenter of the orbit decreases with time. This is not a consequence of dynamical friction, which is not included (i.e., the host halo is an analytical potential), nor of self-friction due to torques from the satellite's own tidally stripped material \citep[][]{Fujii2006, Fellhauer2007} which is negligbly small in our case \citep[see][]{Ogiya2019, Miller2020}. Rather, this shrinking of the orbital extent is simply a consequence of the fact that the host halo is growing (adiabatically) in mass, as described in \autoref{ssec:host_gal}.

The middle panel of \autoref{fig:orbit_mass} shows that the DM mass is rapidly reduced at each pericentric passage where the satellite galaxy feels the strongest tidal force. Prior to the third pericentric passage, there is no evident difference in the DM mass loss history between the two models. This is because the DM mass is preferentially removed from the halo outskirt where the density structure of the two models is almost identical (\autoref{fig:density_profile}). At later times, though, the satellite with a central cusp ($m=0$, dashed line) experience less mass loss than its cored equivalent, simply because the former is more resilient to tides \citep[see e.g.,][]{Penarrubia2010}.
After 10Gyr of evolution, the cored (cuspy) satellite has lost 98.5 (96.7) percent of its initial DM mass. Note that these are lower limits, since in reality dynamical friction (absent in our simulations) would have made the satellite sink further towards the centre of the host halo, where the tidal forces are stronger.

Next, we discuss the evolution of the satellite's stellar mass. Whenever we compute this stellar mass we include the contribution from the GCs. We emphasize, though, that the combined mass of the ten GCs only accounts for five percent of the total stellar mass, and excluding the GCs from the stellar mass therefore has no significant impact on any of our results.

In addition to the bound stellar mass, $\Mbound$, we also determine an estimate of the {\it detectable} stellar mass, $\Mdet$, which is more reminiscent of what an observer would measure. We do so as follows. We project the distribution of star particles onto a plane along a randomly chosen orientation axis, which we register on a 2D rectangular grid with pixels that are $50\pc \times 50\pc$ in size. We convert the projected number density into a surface brightness assuming a mass-to-light ratio of $M/L = 2 (M/L)_{\odot}$ and a distance of $20\Mpc$, similar to what \cite{vanDokkum2018_df2} assumed for DF2. The detectable mass, $\Mdet$, is then defined as coming from those pixels that have a surface brightness exceeding 29\, magn/arcsec$^{2}$, similar to the detection limit of the Dragonfly Telephoto Array used by \citet{vanDokkum2018_df2}, and that are within $10 \kpc$ from the projected centre of the galaxy. For each simulation output we repeat this procedure 100 times, for 100 different random orientation directions, which we use to estimate the error on $\Mdet$. The blue and cyan lines in the lower panel of \autoref{fig:orbit_mass} show the evolution of $\Mbound$ and $\Mdet$, respectively. The error bars on the latter represent the 15-85 percentiles of the distribution obtained from the 100 orientation directions, while the lines show the median. The bound and detectable stellar masses are in good agreement. Both with and without a cusp, the stellar mass of the satellite galaxy is unaffected by the tides until the third pericentric passage ($t \sim -4.5\Gyr$), after which it starts to decline following each subsequent pericentric passage. After $10 \Gyr$ of evolution the detectable stellar mass has declined from $2 \times 10^8 \Msun$ to  $\sim 1.4 \times 10^8 \Msun$ (a reduction of 30 percent) in the case with a cored DM halo, and to $\sim 1.8 \times 10^8\Msun$ (a reduction of 10 percent) in the case of a cuspy DM halo.

In what follows we refer to the {\tt Tide} simulation with a cored DM halo as our reference run.

\subsection{Evolution of galaxy properties}
\label{ssec:galaxy}

\begin{figure*}
    \begin{center}
        \includegraphics[width=0.8\textwidth]{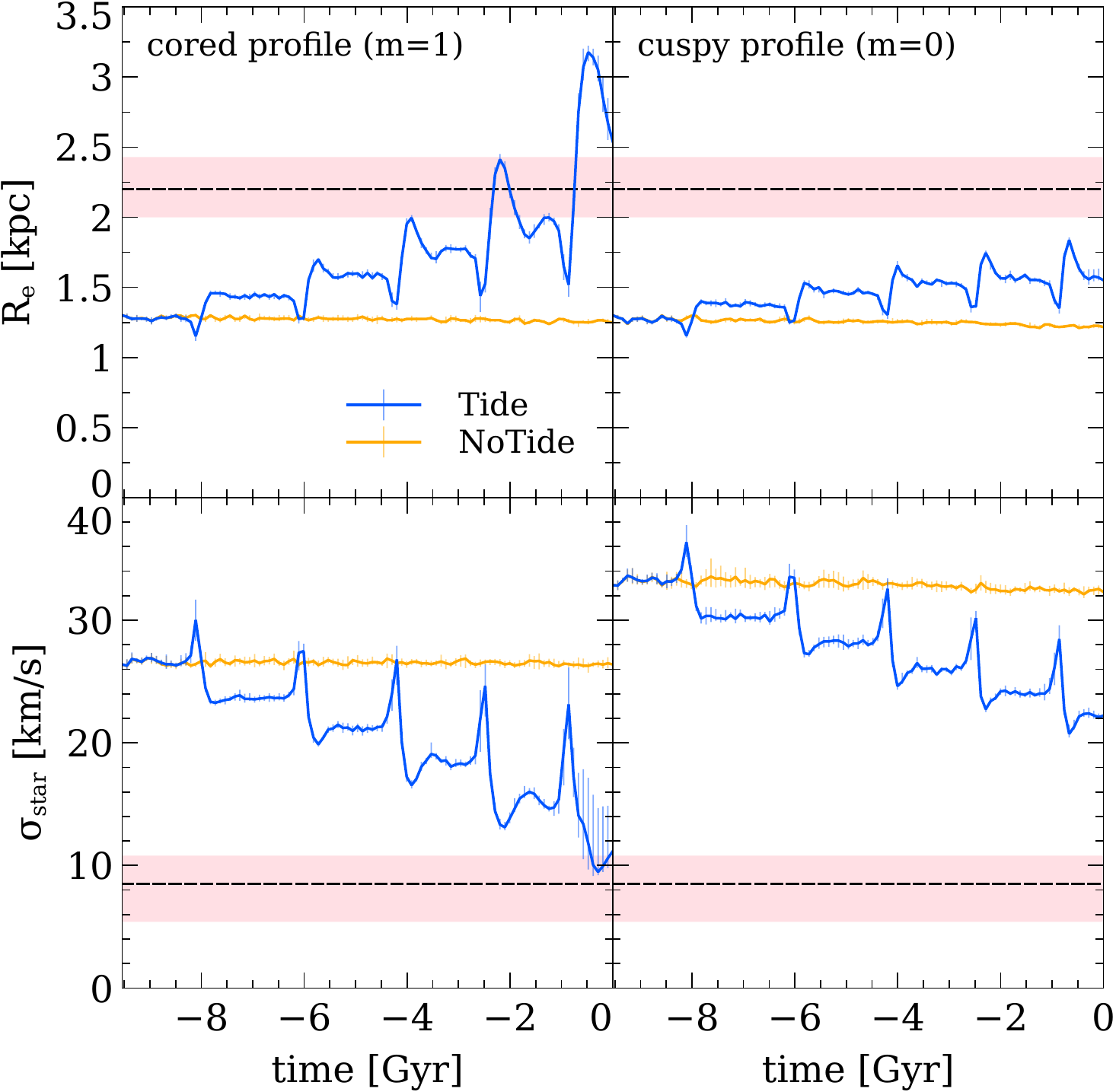}
    \end{center}
    \caption{
        Evolution of galaxy properties. Blue and orange lines are the results from {\tt Tide} and {\tt NoTide} simulations. Simulations shown in the left and right columns adopt the cored DM density profile and the cuspy NFW profile, respectively. {\it Upper Panels:} effective radius, $R\sub{e}$, of the stellar body as function of time. For comparison, the black dashed horizontal line indicates the effective radius of DF2 for an assumed distance of $20 \Mpc$ \citep{vanDokkum2018_df2}. The upper and lower boundaries of the shaded region indicate the effective radius for assumed distances of $D=22.1\Mpc$ \citep{Shen2021_distance} and $18.9\Mpc$ \citep{Cohen2018}, respectively. {\it Lower Panels:} the line-of-sight velocity dispersion of the stars within the effective radius, $\sigma\sub{star}$, as function of time. For comparison, the black dashed horizontal line and the shaded region mark the observed stellar velocity dispersion of DF2 of $\sigma\sub{star}=8.5^{+2.3}_{-3.1} \kms$ \citep[][]{Danieli2019}. Errorbars, which occasionally are too small to see, indicate the 15-85 percentile range obtained using 100 random orientation angles. Note how tidal stirring causes the stellar body of the satellite to expand, and it velocity dispersion to decrease. In the absence of tides, there is no discernible evolution in either $R\sub{e}$ or $\sigma\sub{star}$, indicating that the satellite is dynamically stable in isolation.
    \label{fig:galaxy}}
\end{figure*}

\autoref{fig:galaxy} shows how the tidal forces impact the effective radius $R\sub{e}$ (upper panels) and the line-of-sight velocity dispersion $\sigma\sub{star}$ of the stellar body (lower panels). Here, $R\sub{e}$ is defined as the projected distance from the centre of the galaxy that encloses $\Mdet/2$, and $\sigma\sub{star}$ is the line-of-sight velocity dispersion for the stars within $R\sub{e}$. The lines show the median values obtained from our 100 random viewing angle, while the errorbars mark the 15-85 percentiles. Left- and right-hand panels show the results in the cases where the DM halo of the satellite galaxy has a central core ($m=1)$ and a central cusp ($m=0$). The blue and orange lines show the results obtained from the {\tt Tide} and {\tt NoTide} simulations. Note that the latter reveals no noticeable changes with time, indicating that, in the absence of tidal forces, the system is in dynamical equilibrium. For comparison, the horizontal dashed lines and the shaded regions indicate the $R\sub{e}$ and $\sigma\sub{star}$ of DF2 and their uncertainties.

During each peri-centric passage, which are roughly spaced $2$-$1.5$ Gyr apart\footnote{Due to the growth of the host potential, the orbital period slowly decreases with time.}, the effective radius of the stellar body first undergoes a small decrease, followed by a larger increase, sometimes associated with some oscillation in $R\sub{e}$. The net trend is such that the stellar body `puffs-up' over time. The line-of-sight velocity dispersion reveals a similar, but opposite trend; it first increases slightly, after which it decreases, such that over time the stellar body becomes colder. These are the hallmarks of tidal stripping and impulsive heating. Both are present, and both have a similar impact. 

At peri-center, the tidal radius, defined as the satellite-centric radius where the force from the host halo is equal to that due to the satellite, is smallest. To good approximation, the material outside of this `peri-tidal' radius\footnote{Following \citet{vandenBosch2018a} we refer to the tidal radius at peri-center as the peri-tidal radius.} is stripped off during a short time-interval following peri-centric passage \citep[see e.g.,][and references therein]{Penarrubia2010, Ogiya2019}. Following this mass stripping, the system is out of equilibrium and undergoes re-virialization during which it expands. The energy needed for increasing the (negative) potential energy is taken from the kinetic energy, which explains why the velocity dispersion declines. 

In addition to this purely tidal process, the high-speed pericentric encounter also transfers orbital energy into internal kinetic energy of the satellite system (impulsive heating). Due to the negative heat capacity of a gravitational system \citep[][]{Binney2008}, it responds by transferring kinetic into potential energy, to the extent that the final kinetic energy is lower than the initial value (i.e., the injection of heat has resulted in a colder system). During this re-virialization the kinetic energy is transferred to potential energy, causing the system to expand. This process is often referred to as tidal shocking \citep[][]{Gnedin1999, Gnedin1999_shielding}. 

Which of the two processes dominates, tidal stripping or tidal shocking, depends on the orbit, with tidal shocking become more important for more radial orbits \citep[][]{vandenBosch2018a}. For the purpose of this paper, all that matters is that both processes conspire to puff-up the stellar body of our satellite galaxy, while simultaneously lowering its velocity dispersion. As a consequence, after $\sim 10 \Gyr$ the stellar body in the cored halo has an effective radius and line-of-sight velocity dispersion that are comparable to the values observed in DF2. Although we tuned the orbital circularity to achieve this, the main point is that DF2-like galaxies can be created with tidal interactions starting from a `normal' galaxy.

In the case where the DM halo has a central cusp (right-hand panels), the trend is the same as for a cored halo, but the effect is significantly weaker. Due to the deeper potential well associated with the central cusp, the system is more resilient to tidal effects.

\subsection{Towards Dark Matter Deficiency}
\label{ssec:enclosed_mass}

\begin{figure}
    \begin{center}
        \includegraphics[width=0.4\textwidth]{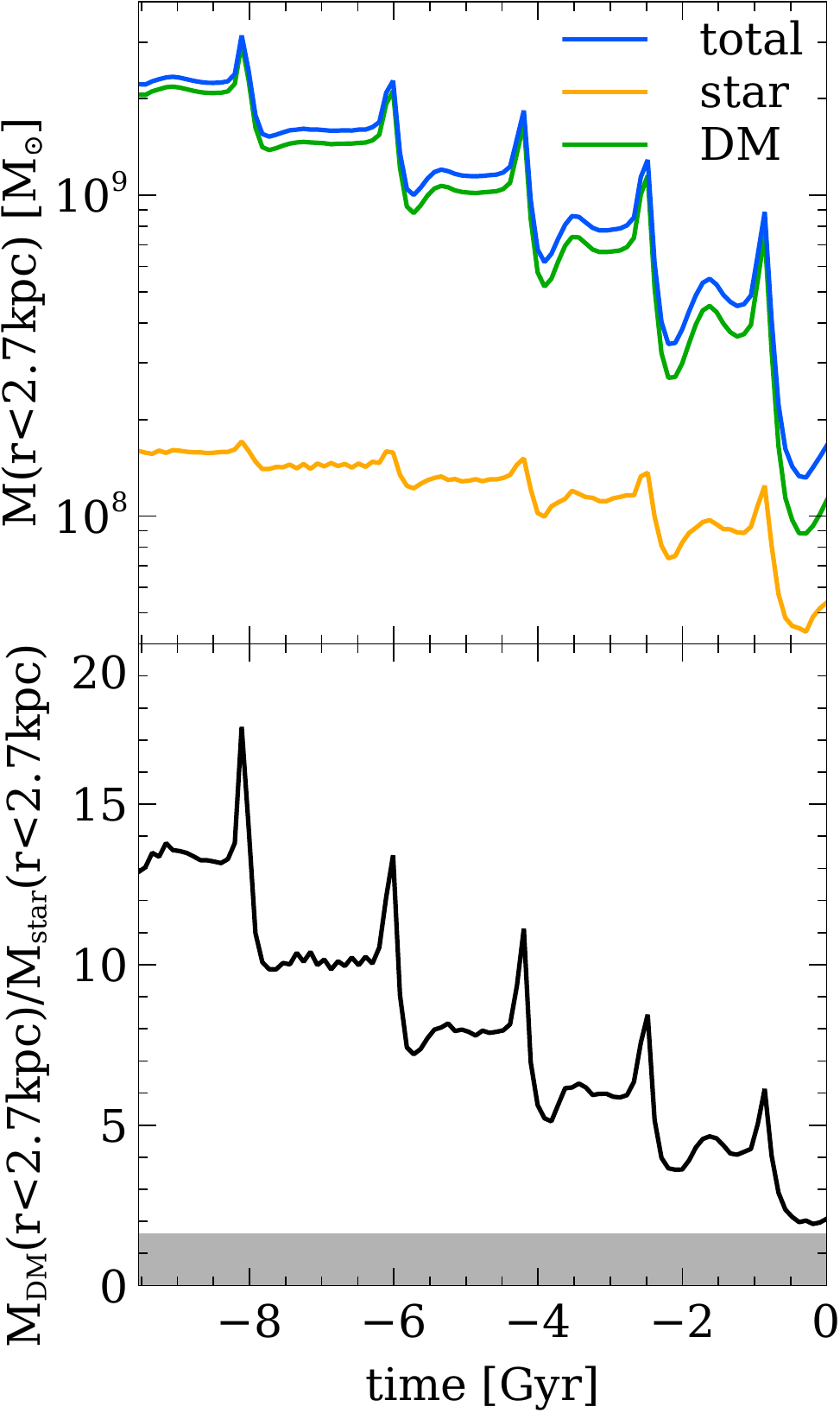}
    \end{center}
    \caption{{\it Upper panel:} Evolution of the enclosed mass within $2.7\kpc$ from the centre of the \nbody satellite galaxy in the reference run. Blue, orange and green lines indicate the total, stellar and DM mass, respectively. {\it Bottom panel:} Ratio of the enclosed DM mass to the enclosed stellar mass. The grey shaded region indicates the 95\% confidence interval on corresponding mass ratio inferred for DF2 by \citet{Danieli2019}. Note how tidal stirring transforms the satellite from a normal, DM dominated galaxy, into a DMDG reminiscent of DF2.
    \label{fig:enclosed_mass}}
\end{figure}

The upper panel of \autoref{fig:enclosed_mass} plots the evolution of the stellar (orange curve) and DM (green curve) masses enclosed within a radius of $2.7\kpc$, which is the  inferred three-dimensional half-light radius of DF2 \citep[][]{Danieli2019}. Note that although the total stellar mass remains constant during the first three pericentric passages (cf., Fig~\ref{fig:orbit_mass}), the stellar mass enclosed within $2.7\kpc$ slightly decreases during each pericentric passage. This is a consequence of tidal stirring. Note also, that the DM mass enclosed within the same radius decreases much more rapidly, which is simply a consequence of the fact that the DM is less tightly bound. 

The lower panel plots the ratio of the enclosed DM mass to the enclosed stellar mass. The gray-shaded region indicates the 95\% confidence region on this mass ratio using the constraints on the enclosed stellar mass and the total enclosed dynamical mass from \cite{Danieli2019}. The satellite system starts out being completely dominated by DM, with $M\sub{DM}(r<2.7\kpc)$ about 13 times higher than the stellar mass enclosed within the same radius. After each pericentric passage, though, the DM dominance is reduced. At the end of our simulation, after five pericentric passages, the DM mass enclosed with $2.7\kpc$ is only 2 times that of the enclosed stellar mass; the satellite has become a DMDG, with properties that are similar to what has been inferred for DF2.

\subsection{Evolution of the GC population}
\label{ssec:gc}

\begin{figure*}
    \begin{center}
        \includegraphics[width=0.8\textwidth]{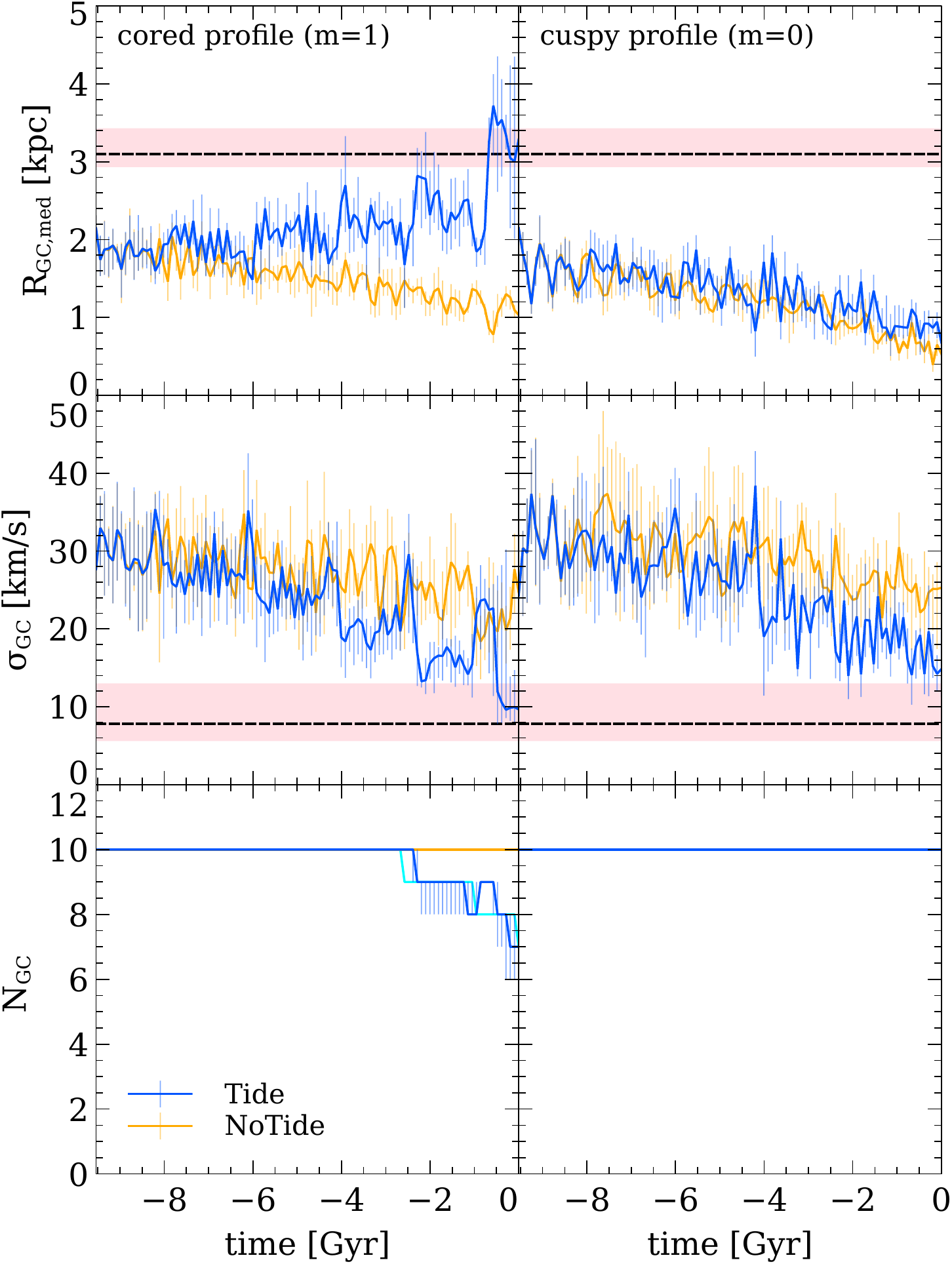}
    \end{center}
    \caption{Evolution of the GC population. Blue and orange lines show the results from {\tt Tide} and {\tt NoTide} simulations, while left- and right-hand columns show the simulation results that adopt the cored and cuspy DM halo, respectively. {\it Top panel:} median projected radius, $R\sub{GC,med}$, of the GCs from the centre of the satellite galaxy, indicating the spatial extent of the GC distribution. For comparison, the black dashed horizontal line indicates the corresponding radius in DF2 assuming a distance of $20\Mpc$ \citep[see][]{vanDokkum2018_df2}, while the upper and lower boundaries of the shaded region mark the corresponding radii using the distance measurements of \citet{Shen2021_distance} and \citet{Cohen2018}, respectively. {\it Middle panel:} line-of-sight velocity dispersion among the GCs, $\sigma\sub{GC}$. For comparison, the black dashed horizontal line and shaded region mark the observational measurements, $\sigma\sub{GC} = 7.8^{+5.2}_{-2.2}\kms$, for DF2 \citep{vanDokkum2018_correction}. {\it Bottom panel:} number of GCs with a projected separation $R < 10\kpc$. For comparison, the number of GCs that are actually bound to the \nbody system in the {\tt Tide} and {\tt NoTide} simulations are shown as cyan and yellow lines, which is in good agreement with those at $R < 10 \kpc$. Note that in the analyses of the top and middle panels, only GCs with $R < 10\kpc$ are considered. As always, errorbars indicate the 15-85 percentile range obtained using 100 random orientation angles, while the lines show the median. Note how, in the absence of tides, the GCs sink to the centre due to dynamical friction, resulting in a reduction of both $R\sub{GC,med}$ and $\sigma\sub{GC}$, as discussed in the text. In the case where the halo is cored, tidal stirring is strong enough to overcome this mass-segregation, causing the distribution of GCs to expand following each peri-centric passages.
    \label{fig:gc}}
\end{figure*}

The upper four panels of \autoref{fig:gc} are similar to \autoref{fig:galaxy}, except that this time they depict the evolution in the median, projected radius, $R\sub{GC, med}$, and the line-of-sight velocity dispersion, $\sigma\sub{GC}$, of the GCs rather than the stars. As with the stars, these are obtained using 100 random orientation angles per simulation output. For each projection, only the GCs with a projected radius $R < 10\kpc$ are included in the measurement; GCs that are further away from the projected centre of the galaxy are considered stripped off\footnote{For comparison, the GC in DF2 with the largest projected galacto-centric radius has $R=7.6\kpc$ \citep{vanDokkum2018_df2}.}. The physical validity of this oversimplified `stripping criterion' is discussed below.

In the case where the halo of the satellite galaxy is cored (left-hand panels), the results are similar to what is seen for the stars: the tidal effects cause the population of GCs to expand following each pericentric passage, while their line-of-sight velocity dispersion decreases. After $10\Gyr$ the results are comparable to what is observed for DF2 (indicated by the horizontal dashed lines plus shaded regions). If the halo is cuspy (right-hand panels), the evolution of $R\sub{GC, med}$ is profoundly different. Rather than increasing over time, it actually decreases in a fashion that is similar to what is seen in the {\tt NoTide} simulations (orange curves). As discussed in \S~\ref{sssec:heatingcooling} below, this is due to dynamical friction.

The bottom panels of \autoref{fig:gc} show the number of GCs with $R<10\kpc$ in the {\tt Tide} (blue lines) and {\tt NoTide} (orange lines) simulations, respectively. For comparison, the cyan and yellow lines indicate the corresponding numbers of GCs that are gravitationally bound to the satellite galaxy. Note that the number of GCs with $R < 10 \kpc$ is a reasonable `proxy' for bound GCs. In the {\tt NoTide} simulations the number of GCs that remain bound and/or within a projected separation of $10\kpc$, remains constant at 10. And the same is true in the {\tt Tide} simulations with a cuspy halo. However, when the satellite galaxy has a cored halo and tides are present, GCs occasionally are stripped off. After $10 \Gyr$ between 60 and 80 percent of the original GCs remain bound and/or within a projected separation of $10\kpc$. Note, though, that stripping of the GCs does not start until after the third pericentric passage.

Interestingly, the empirical relation between the number of GCs and halo virial mass of \cite{Burkert2020} suggests that our satellite galaxy, which has an infall mass of  $M\sub{200} = 6 \times 10^{10} \Msun$, should have 12 GCs on average. Hence, the fact that DF2 is observed to have 10 GCs within  $R=7.6\kpc$ \citep{vanDokkum2018_df2} is perfectly consistent with DF2 being a tidal remnant of what used to be a perfectly normal galaxy for a halo of that infall mass.

Since the initial position and velocity vectors of the DM particles, stars and GCs of the satellite galaxy are all drawn randomly from their respective distributions, there is non-zero realisation-to-realisation variance. This could potentially be very significant for the GCs because of their small number. In order to test the impact of this `realization-variance' we run nine additional {\tt Tide} simulations with different random realizations for the phase-space coordinates of the \nbody satellite system.\footnote{Studying realization variance is also useful for revealing potential numerical instabilities, as in the case of artificial disruption \citep[e.g.,][]{vandenBosch2018b}.} We find that the realization-to-realization variance in the evolution of $R\sub{e}$, $\sigma\sub{star}$, $R\sub{GC,med}$, $\sigma\sub{GC}$ and $N\sub{R < 10\,{\rm kpc}}$ are all very comparable to the variance arising from the random orientation angles. Hence, the errorbars in Figs.~\ref{fig:galaxy} and~\ref{fig:gc} roughly reflect the uncertainties due to realization-variance. As expected, these are significantly larger for the GCs than for the stars.

\subsubsection{Mixing and Dynamical Segregation}
\label{sssec:heatingcooling}

The evolution of $R\sub{GC,med}$ and $\sigma\sub{GC}$ reflects a competition between processes that cause dynamical segregation of the GCs with respect to the stars and DM particles and processes that tend to undo such segregation. 

\begin{figure}
    \begin{center}
        \includegraphics[width=0.4\textwidth]{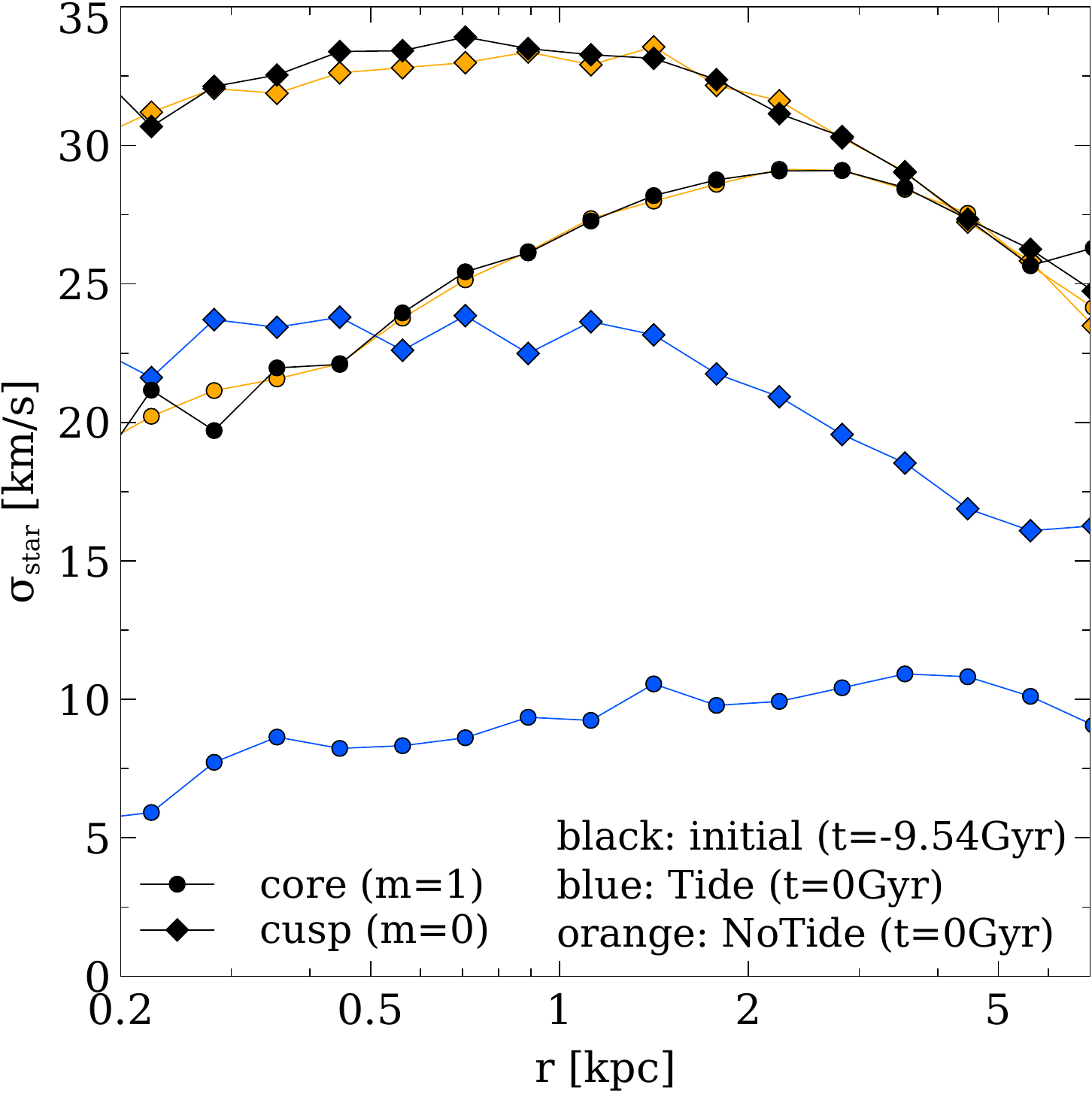}
    \end{center}
    \caption{Velocity dispersion profile of stars. Black lines represent the initial configuration, while blue and orange lines are the snapshot at the present time in the {\tt Tide} and {\tt NoTide} simulations, respectivey. Lines with circles (diamonds) indicate the cases where the DM halo of the satellite galaxy has a cored (cuspy) density profile. In the absence of tides (orange lines), the profiles do not evolve, once again indicating that the satellite is stable in isolation. When the satellite is subjected to tidal stripping, the stellar velocity dispersion decreases with time, but the shape of the radial profile remains similar to its initial shape.
    \label{fig:vel_disp_profile}}
\end{figure}

In the absence of tides, the GCs only experience dynamical friction, which causes the GCs to undergo mass segregation with respect to the stars. Note that in doing so, $\sigma\sub{GC}$ decreases with time. This is simply a consequence of the detailed potential of the satellite galaxy. \autoref{fig:vel_disp_profile} shows the velocity dispersion profiles of stars in the {\tt Tide} and {\tt NoTide} simulations, both initially (black) and at the present time (i.e., after $10\Gyr$ of evolution). In the absence of tides, the stellar velocity dispersion profile remains invariant, indicating that the dynamical friction experienced by the GCs does not significantly boost the velocity dispersion of the stars. This is simply because the total mass of the GCs is small compared to the total stellar mass of the system. In other words, the GCs can be treated as a tracer population in quasi-equilibrium (i.e., the dynamical friction time scale is long compared to the dynamical time). Consequently, its velocity dispersion should resemble that of stars with a similar radial distribution. Since $\sigma\sub{star}$ declines towards the centre, so does the velocity dispersion among the GCs as they slowly sink towards the centre due to dynamical friction.

In the presence of tides there are two processes that act against mass segregation. First of all, re-virialization in response to tidal stripping is a form of violent relaxation, which is independent of the mass of the constituent particles \citep[][]{Lynden-Bell1967}. The changes in specific energy that result from re-virialization only depend on orbital phase, which are randomly distributed for both the stars and the GCs. Hence, re-virialization has a tendency to mix the GCs and stars, and thus to undo any mass segregation that might have occurred as a consequence of dynamical friction. Secondly, the impulsive shock resulting from the high-speed peri-centric passsage imparts the stars and GCs with a velocity impulse that is independent of their mass, but proportional to their satellite-centric distance squared \citep[][]{Spitzer1958, Gnedin1999, Banik2021_tidal}. Hence, objects that are further from the centre of the satellite receive more kinetic energy. During the re-virialization that ensues twice the total kinetic energy that has been imparted to the satellite galaxy is transferred to potential energy \citep[][]{Binney2008}. Despite a fair amount of mixing that occurs during this re-virialization, the particles that gained more kinetic energy will tend to move out further than their counter-parts for whom the velocity impulse was weaker. Consequently, if the GCs start out being distributed differently from the stars, tidal shocking tends to enhance this difference. It causes a segregation, not by mass but by specific binding energy. Hence, if the GCs are less centrally concentrated than the stars prior to the tidal shock, the latter will act in the opposite direction of mass segregation induced by dynamical friction. 

Our simulations indicate that in all cases considered dynamical friction dominates in the sense that the GCs are always found to undergo net mass segregation. In the absence of tidal forces this is obvious, but even when tides are present this is the case. If the satellite has a central cusp, $R\sub{GC,med}$ decreases while the effective radius of the stars increases. Clearly, re-virialization and tidal shocking are not strong enough to overcome the mass segregation induced by dynamical friction. And even in the case where the satellite has a central core, and dynamical friction is significantly weaker, there is still a net mass segregation: while the effective radius of the stars increases by a factor two (from $1.25\kpc$ to $2.5\kpc$) over the $10\Gyr$ duration of the simulation, the median radius of the GC population, $R\sub{GC,med}$, only increases by a factor 1.5 (from $2\kpc$ to $3\kpc$) over the same time interval. 

\subsubsection{Evolution of individual GCs}
\label{sssec:individual_GCs}

\begin{figure}
    \begin{center}
        \includegraphics[width=0.4\textwidth]{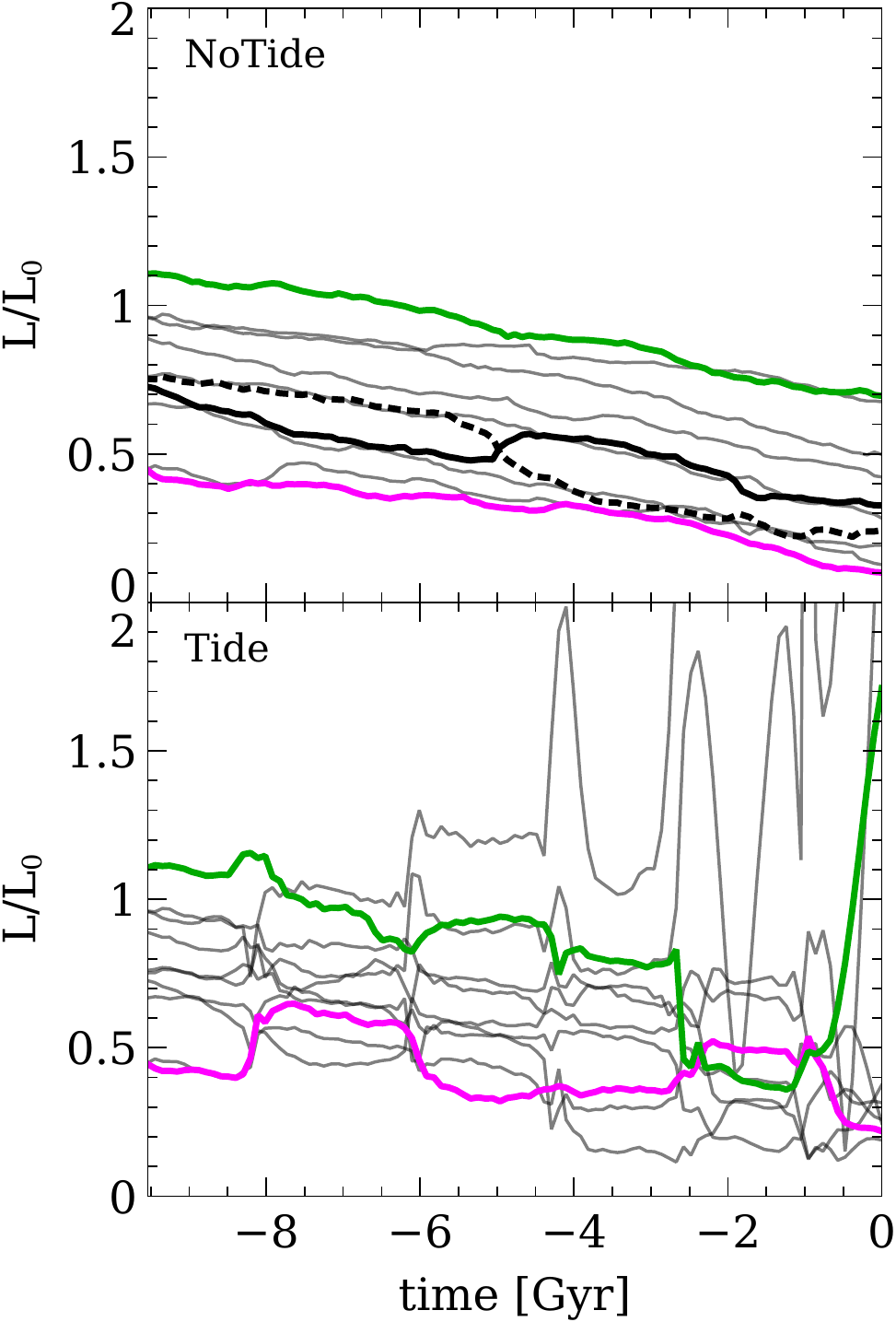}
    \end{center}
    \caption{The evolution in orbital angular momentum for the ten individual GC particles in a {\tt NoTide} simulation (upper panel) and in the reference {\tt Tide} run (lower panel). The initial conditions in both cases are identical. As a function of time, we plot the norm of the specific orbital angular momentum of the GCs with respect to the centre of the satellite galaxy, $L$, scaled by the specific orbital angular momentum of a circular orbit at $r=r\sub{GC}$, $L\sub{0}$. To guide the eye, the green and magenta curves correspond to the GCs that start out with the largest and smallest orbital angular momentum, respectively. In the absence of tides, $L/L\sub{0}$ steadily decreases due to dynamical friction, and except for occasional GC-GC scattering (an example is indicated with the black solid and dotted lines), the rank-order of $L/L\sub{0}$ is reasonably well preserved. In the presence of tidal stirring, though, this rank-order is scrambled following each peri-centric passage, and the variance in $L/L\sub{0}$ increases with time.
    \label{fig:individual_GCs_L}}
\end{figure}

\autoref{fig:individual_GCs_L} shows how the orbital angular momentum, $L$, of individual GCs evolve in a {\tt NoTide} simulation (upper panel) and the reference {\tt Tide} run (lower panel). The initial phase-space coordinates for the stars and GCs in both simulations are identical. Green and magenta curves correspond to the GCs with the largest and smallest initial angular momenta, respectively. 

In the absence of tides, the angular momenta of the GCs steadily decrease with time as a consequence of dynamical friction. Occasionally, two-body interactions (scattering) among two GCs cause a mutual exchange of angular momentum, as in the case of the GCs indicated with the thick solid and dotted black curves, which undergo mutual scattering around $t \sim -5\Gyr$. Occasionally, these scattering events can significantly boost the orbital angular momentum of the GCs, thereby significantly delaying the time scale over which they sink to the centre of the satellite galaxy \citep[cf.][]{DuttaChowdhury2019}. The overall impact of GC-GC scattering is fairly limited though, such that on average the GCs maintain their angular momentum rank-order. 

This is very different in the {\tt Tide} simulations, where tidal stirring causes rapid changes in the orbital angular momenta of individual GCs, especially shortly after the satellite galaxy has a peri-centric passage within the host halo. Whether a GC gains or looses angular momentum during such a peri-centric passsage depends on its orbital phase within the satellite at that moment in time. Since these are different for each of the GCs, they all respond differently, which causes the angular momentum rank-ordering to be scrambled following each peri-centric passage. For instance, note how the green and magenta lines cross, indicating that the angular momentum of the GC that initially had the highest value of $L$ is (temporarily) surpassed by the GC that started out with the lowest $L$. Overall, tidal stirring cause a significant increase in the variance of the orbital angular momenta among the GCs.

\section{Internal structure of the satellite galaxy}
\label{sec:internal_structure}

Having demonstrated that tidal forces can transform a normal (average) galaxy into a DMDG with sizes and velocity dispersions that are similar to those of DF2, we now examine how the detailed internal structure of the satellite galaxy evolves as it is subjected to the tides. In particular, we examine how tidal stirring impacts the radial profile of the satellite as well as metallicity gradients. Since stirring causes a scrambling of rank-orders, the expectation is that any pre-existing radial gradients in metallicity, stellar age, or any other property, will be diminished over time.

\subsection{Surface brightness profile}
\label{ssec:surface_brightness}

\begin{figure}
    \begin{center}
        \includegraphics[width=0.4\textwidth]{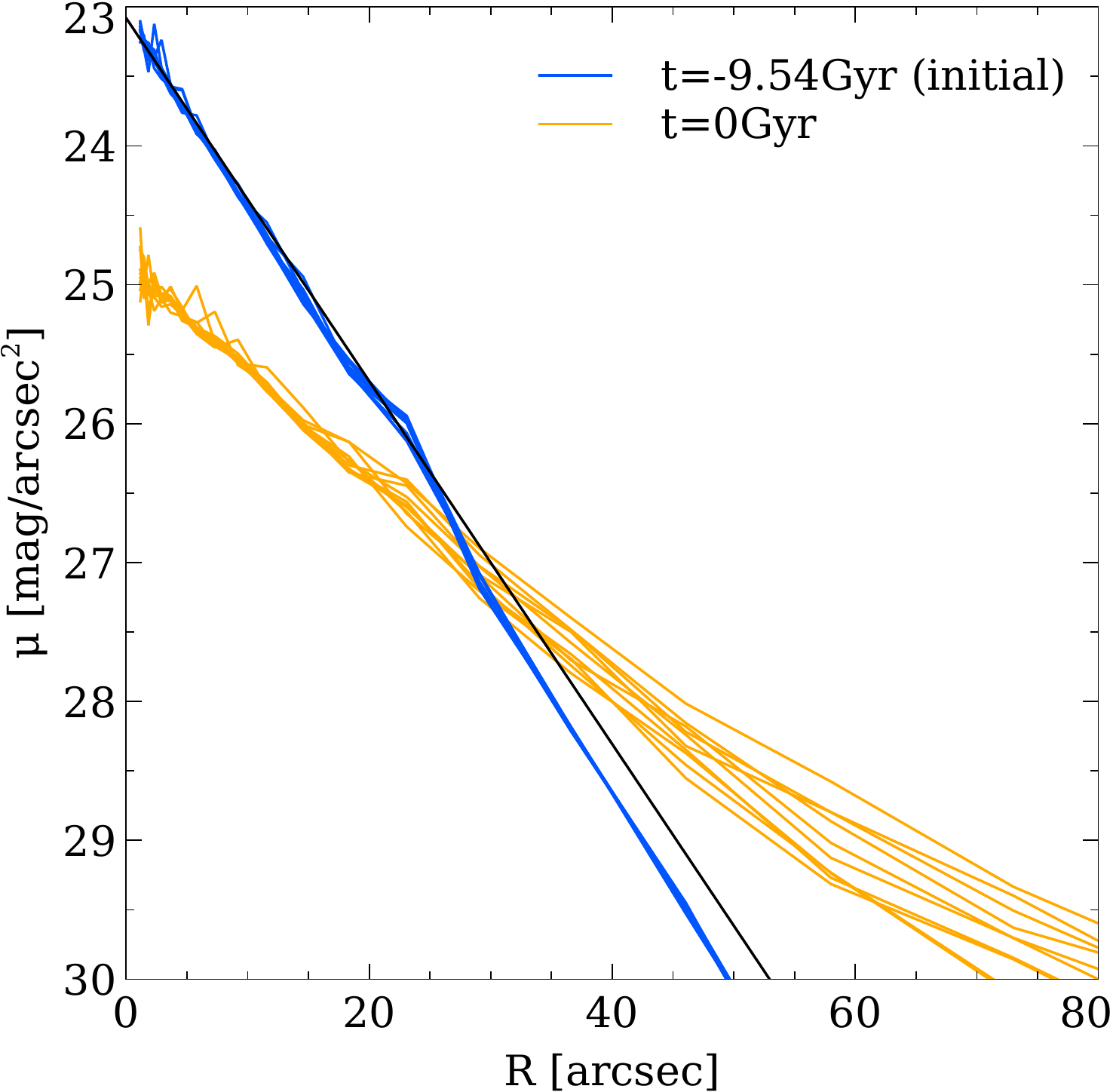}
    \end{center}
    \caption{Stellar surface brightness profiles. The solid black curve indicates the analytical $n=1$ \sersic profile used to initialize the positions of the stellar particles. The blue lines show the initial surface brightness profile along 10 random orientation angles, obtained by converting the projected mass density adopting a stellar mass-to-light ratio of $M/L = 2(\Msun/\Lsun)$ and a distance of $20\Mpc$. The orange lines show the stellar surface brightness profiles (again using 10 random orientation angles) of the satellite galaxy in our reference {\tt Tide} run after $10\Gyr$ of evolution. Note how tidal stirring has caused a decrease in the central surface brightness by about 2 magnitudes, and an overall expansion, while maintaining a roughly exponential surface brightness profile throughout.
    \label{fig:surface_brightness}}
\end{figure}

The blue curves in \autoref{fig:surface_brightness} show the initial surface brightness profile of our \nbody satellite galaxy, with different curves corresponding to different, random projections. For comparison, the solid black curve shows the surface brightness profile of the $n=1$ \sersic profile used to initialize the particle positions as described in \S\ref{sssec:stellar_spheroid_satellite}. Orange curves show the surface brightness profile of the \nbody satellite system at the end of the simulation. Over time, tidal stripping and shocking have caused the stellar system to puff-up, such that the final system is significantly more extended than initially. Note, though, that the tidal forces have not only affected the outskirts. In fact, due to tidal stirring the central surface brightness has decreased by a factor of 6.3, from 23 mag/arcsec$^{2}$ to 25 mag/arcsec$^{2}$. Interestingly, despite the drastic changes to the satellite's surface brightness profile, it has roughly remained exponential. Whether this is a mere coincidence, or whether there is a physical reason for this shape-invariance, is a question that we leave for future exploration. 

\subsection{Radial mixing}
\label{ssec:redistribution_stars}

\begin{figure}
    \begin{center}
        \includegraphics[width=0.4\textwidth]{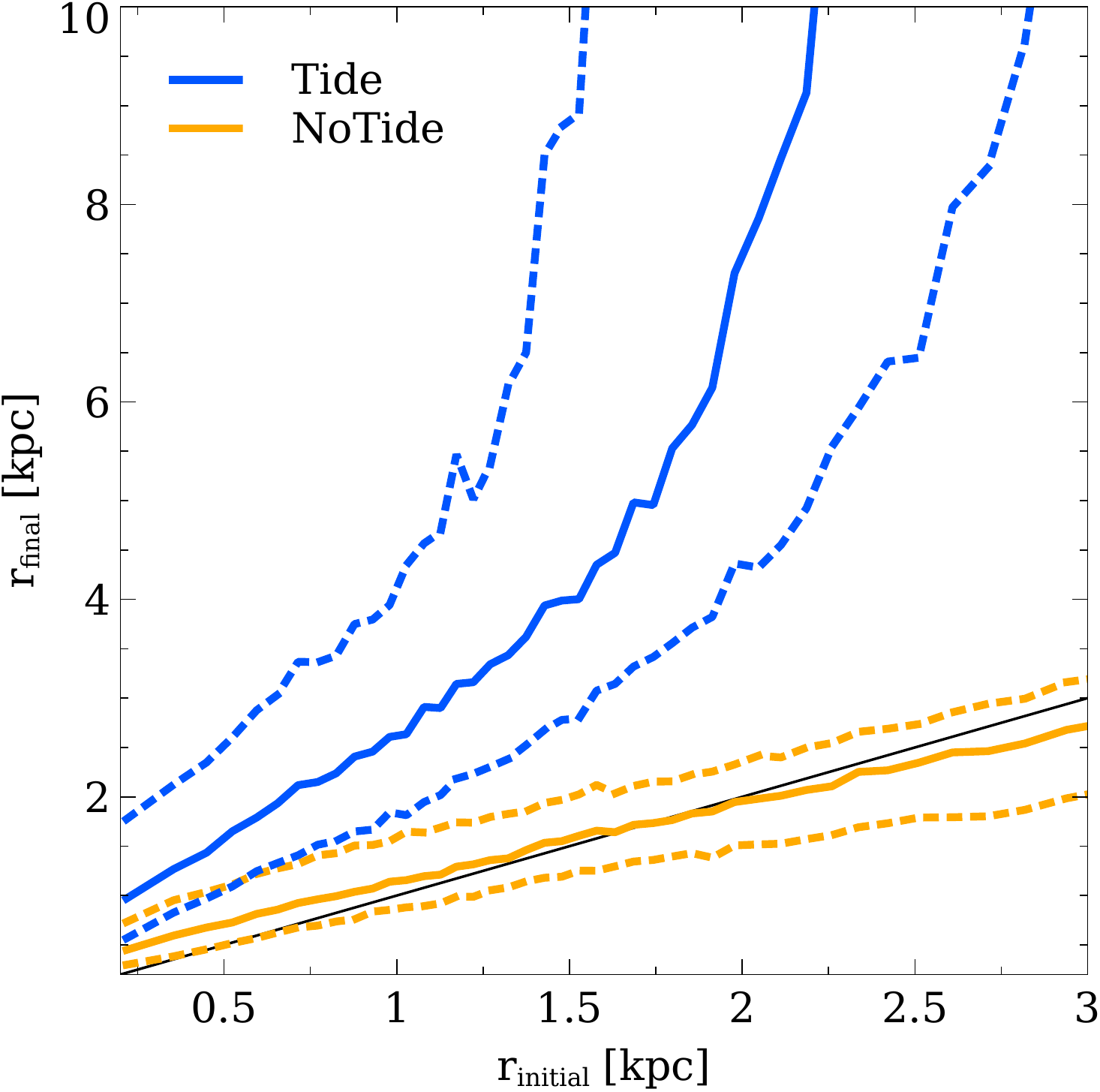}    
\end{center}
    \caption{Distribution of stars in the $r\sub{initial}-r\sub{final}$ plane, where $r\sub{initial}$ and $r\sub{final}$ are the distance from the centre of the \nbody system to each star particle at the beginning of the simulation and at the present time, respectively. Blue and orange lines show the results from the reference run and the {\tt NoTide} simulation of the same initial \nbody configuration. The solid line and the lower and upper dotted lines indicate the median and the 25 and 75 percentiles of the distribution. The black line marks the line of identity, i.e.,  $r\sub{final} = r\sub{initial}$, and is shown for comparison. Note how tidal stirring causes an overall expansion, and, especially at larger radii, mixing that increases the variance in $r\sub{final}$ for a given $r\sub{initial}$. The small scatter and slight deviation from identity in the {\tt NoTide} case is a consequence of the fact that stars are on non-circular orbits, as discussed in the text.
    \label{fig:rirf_distribution}}
\end{figure}

As eluded to above, tidal stirring causes a scrambling of the rank order distributions in binding energy and orbital angular momentum. Hence, it will tend to `diffuse' away any pre-existing radial gradients in for example stellar age and/or metallicity. In order to gauge the extent of this radial mixing, \autoref{fig:rirf_distribution} plots the final radii, $r\sub{final}$, of stellar particles versus their initial radii, $r\sub{intial}$. Solid curves indicate the medians, while the dashed curves indicate the 25 and 75 percentiles. The solid, black curve marks the line of equality, i.e., $r\sub{final} = r\sub{initial}$, and is shown for comparison. In the absence of tides (orange curves), the median $r\sub{final}$ is in reasonable agreement with the initial radius. There is some indication that particles that start out with $r\sub{initial} \lta 2 \kpc$ on average end up a little but further out, while the opposite holds for particles that start out with $r\sub{initial} \gta 2 \kpc$. This does not imply that there is a net evolution in the surface brightness distribution of the satellite. Rather, it merely reflects that in an equilibrium configuration, the particles maintain their rank-order distribution in binding-energy, not in radius. Particles at a given initial radius span a range in binding energies. At a later time, phase-mixing assures that particles that all shared the same initial radius are now spread out in radius, which explains the non-zero spread in $r\sub{final}$ for a given $r\sub{intial}$. The fact that the median deviates from the line of equality arises from the fact that at large radii one is more likely to find particles that are near their orbital apo-center. At later times, on average, these particles will be at a smaller radius than initially. Similary, at small radii, one is more likely to find particles that are near their orbital peri-center, and those particles are more likely to be located at a larger radii at some later point in time\footnote{Note that the typical apo-to-pericenter ratio for orbits in isotropic systems is $\sim 4$ \citep[][]{vdbosch.etal.99}.}.

The blue lines in \autoref{fig:rirf_distribution} show the results for our reference run with tides. The majority of particles now have $r\sub{final} > r\sub{initial}$, which indicates once more that the stellar body of the satellite galaxy is puffing up over time. Note that both the ratio $r\sub{final}/r\sub{initial}$ and the variance in $r\sub{final}$ increase with increasing $r\sub{initial}$. These are the hallmarks of tidal stirring, which has a larger effect at larger radii (i.e., the velocity impulse due to high-speed pericentric passages is $\propto r^2$), and which scrambles the rank-ordering in binding energy and angular momentum due to the violent relaxation associated with re-virialization.

\begin{figure}
    \begin{center}
        \includegraphics[width=0.4\textwidth]{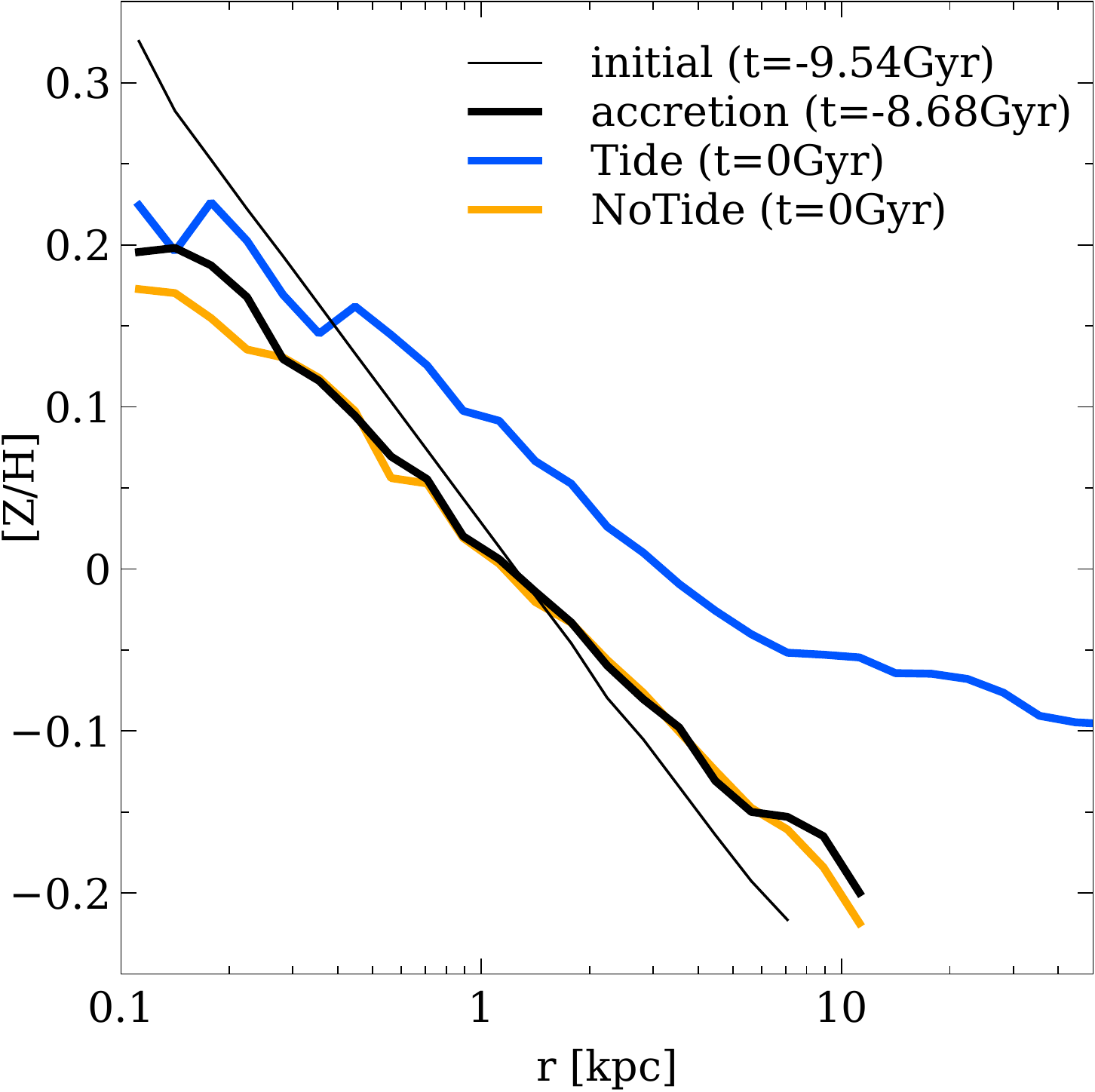}
    \end{center}
    \caption{Metallicity profile. The thin line indicates the initial radial metallicity profile (equation~(\ref{Zprof}) used to assign stars a metallicity based purely on their initial radius, $r\sub{initial}$. The thick black and orange lines shows the metallicity profile in the {\tt NoTide} simulations after $\sim 1\Gyr$ (when the satellite crosses the virial radius of the host for the first time) and $10\Gyr$, respectively. The fact that stars are not on circular orbits causes a quick ($<1\Gyr$ `diffusion' of metallicity which slightly reduces the slope $\rmd [Z/H]/\rmd\log r$, but there is no subsequent, late-time evolution since the binding energies of the star particles are conserved. In the presence of tides (blue line), though, tidal stirring causes a strong reduction of the metallicity gradient at large radii, an observational prediction that can be used to test whether systems like DF2 and DF4 have indeed been subjected to strong tidal stirring. At small radii, the metallicity is slightly boosted, which is a consequence of the fact that tidal forces preferentially strip the lower metallicity, less bound stars. 
    \label{fig:metallicity}}
\end{figure}
 
\autoref{fig:metallicity} illustrates how this affects a radial metallicity gradient. Here we assign each star particle in our simulations a metallicity based purely on its initial radius, $r\sub{initial}$. Guided by the observed metallicity gradients in galaxies of stellar mass $\Mbound \sim 2 \times 10^8 \Msun$ \citep{Peletier2012}, we assign each star a metallicity
\begin{equation}\label{Zprof}
[Z/H] = -0.3 \log[r\sub{initial}/R\sub{e,i}] \,,
\end{equation}
with $R\sub{e,i}=1.25\kpc$ the initial effective radius of the model galaxy. The resulting metallicity as function of radius is shown as the thin black line in \autoref{fig:metallicity}. The thick black line shows the median metallicity as function of radius after $1\Gyr$ of evolution in the {\tt NoTide} simulation. Note that the slope $\rmd [Z/H]/\rmd\log r$ has become somewhat shallower ($-0.2$ as compared to the initial $-0.3$), which is simply a manifestation once more of the fact that, unlike binding energy, galactocentric radius is not a conserved quantity. The orange curve shows the results in the same {\tt NoTide} simulation, but now after $10\Gyr$ of evolution. As one can see, in the absence of tides there is no long-term evolution in the metallicity gradient. The only evolution is a rapid ($< 1 \Gyr$) adjustment to account for the fact that orbits span a range of radii.

The blue line shows the median metallicity as function of radius after $10\Gyr$ of evolution in our reference {\tt Tide} simulation. In the inner region the slope is similar to that in the {\tt NoTide} simulation, but note that the median metallicity has increased. This is a consequence of the fact that the least bound stars have the lowest metallicities and are the stars that are most easily stripped off, or `dispersed' to larger radii as a consequence of the tidal stirring. In other words, the median metallicity at small radii increases because the low-metallicity tail is moved to larger radii. At larger radii ($r \gta 5\kpc$), the metallicity gradient $\rmd [Z/H]/\rmd\log r$ becomes very shallow, which is simply a manifestation of that fact that tidal stirring causes larger mixing at larger radii (cf., Fig.~\ref{fig:rirf_distribution}). Hence, if DMDGs like DF2 have indeed been shaped by strong tidal forces, we expect that they should have very shallow metallicity gradients in their outskirts, something that can in principle be tested observationally.

\section{Summary}
\label{sec:summary}

The recent discovery of two DMDGs, DF2 and DF4, with abnormal GC populations has sparked a vigorous debate as to their origin. Although tidal dwarfs are a natural explanation for DMDGs, in the case of DF2 and DF4 such a formation scenario does not (naturally) explain their large GC systems, nor the fact that DF2 and DF4 seem to have metallicities that are consistent with the typical mass-metallicity relation. An alternative explanation for their origin, proposed by \cite{Ogiya2018} and \cite{Nusser.20}, is that DF2 and DF4 are remnants of fairly extreme tidal mass loss. In particular, using numerical simulations \cite{Ogiya2018} has shown that tidal stripping can transform a fairly normal galaxy into a DMDG with properties reminiscent of DF2, as long as the progenitor halo had a central density core. 

An important question is whether such a stripping scenario can explain the fact that both DF2 and DF4 have large specific frequencies of GCs (i.e., a large number of GCs per unit stellar mass). Tidal stripping and tidal shocking both have a tendency to preferentially remove matter from the outskirts. Hence, if the population of GCs is more extended than that of the stars, tidal processes will tend to {\it lower} the specific frequency of GCs. However, GCs are also subject to dynamical friction, which causes them to migrate inwards, potentially resulting in the formation of a nuclear star cluster\footnote{Unless core-stalling or GC-GC scattering prevents the GCs from reaching the centre \citep[see][]{DuttaChowdhury2019}}. In addition, re-virialization in response to tidal stripping and shocking tends to cause mixing, and thus to undo any mass segregation. In this paper we used idealized numerical simulations to study how all these processes act together, and to see whether, starting from a normal galaxy, tidal stripping can produce a system that is similar to DF2 and DF4 in both the spatial distribution and kinematics of both the stars and the GC population.

Our simulations model the tidal evolution of an \nbody satellite galaxy (i.e., the DF2 progenitor), consisting of a stellar spheroid plus a population of GCs embedded in a DM halo, as it orbits the time-evolving analytical potential of its host galaxy (i.e., NGC~1052). The initial satellite galaxy, prior to infall, is modeled as a `normal galaxy', having a typical stellar mass-to-halo mass ratio, a typical size for its initial stellar mass, and a typical GC population, both in terms of the number of GCs per unit halo mass and in terms of their spatial extent. Its DM halo is modelled to have either a central NFW-like cusp, or a constant density core. The analytical host halo is modelled as a NFW halo, which slowly (adiabatically) grows its mass over time in accordance with the average mass assembly history of halos of that mass. In order to identify the impact of the tidal forces, we compared the results from these {\tt Tide} simulations, with {\tt NoTide} simulations in which the \nbody satellite is evolved in isolation.

As the \nbody satellite orbits its host halo it experiences tidal forces that strip matter from its outskirts. Stripping is dominated during peri-centric passage, which is also when a tidal shock injects energy into the system. The subsequent re-virialization causes the \nbody system to puff up and its velocity dispersion to decrease. We tune the orbit such that, in the case where the initial halo of the \nbody satellite has a central core, the final stellar remnant has an effective radius that is reminiscent of that of DF2. The resulting orbit has a peri-centre during first infall that corresponds to the 8.3 percentile of the distribution of peri-centres for first-infall orbits of DM substructures. After $10\Gyr$ of evolution, which corresponds to 5 peri-centric passages, the DM halo of the \nbody satellite has lost 98.5 percent of its original mass. The stellar body, on the other hand, only lost about 30 percent of its original mass, such that the system has evolved from a normal DM dominated galaxy to a DMDG. In particular, the DM mass fraction inside the central $2.7\kpc$ (roughly the half-mass radius of DF2) has decreased from 13 to 2, while the stellar velocity dispersion has decreased from $\sim 26\kms$ to $\sim 12\kms$. These properties are similar to what has been observed in DF2. 

In the case where the initial halo of the \nbody satellite is cuspy, the tidal effects are significantly weaker, and after $10\Gyr$ the tidal remnant is still too compact, and with a velocity dispersion that is too high, to be considered a viable DF2 look-a-like. It may still be possible to create a system that resembles DF2 by placing it on a more eccentric orbit, but we suspect that this requires an extremely radial orbit, with a peri-centre that would lie well inside the stellar body of NGC~1052. We have not explored this option, though, and instead mainly focus on the results obtained with the initially cored satellite. We emphasise that the initial DM halo mass of our \nbody satellite is $6 \times 10^{10}\Msun$, which is exactly the mass scale where stochastic feedback from supernova is believed to be efficient in creating central halo cores \citep[e.g.,][]{DiCintio2014}.
 
If we initialise the distribution of GCs such that their effective radius in projection is $R\sub{e,GC} \sim 2\kpc$, which is consistent with, but at the low end of the observed distribution of $R\sub{e,GC}$-values for galaxies with a stellar mass similar to that of our initial satellite \citep[cf.][]{Forbes.17}, we find that the tidal effects yield a final GC distribution, after $10\Gyr$ of evolution that is consistent with what has been observed in DF2. In particular, the median projected radius has increased from $2\kpc$ initially to $\sim 3\kpc$, while the line-of-sight velocity dispersion among the GCs has decreased from $\sim 30\kms$ initially to $\sim 10\kms$. Note that the increase in the median projected radius is less pronounced than for the stellar body, which is a consequence of dynamical friction, which causes ongoing mass segregation which is countered by the re-virialization following each peri-centric passage. On average, only about 20 percent of the GCs are stripped off. Hence, given that a halo with an infall mass of $M\sub{200} = 6 \times 10^{10} \Msun$ hosts on average 12 GCs \citep[see][]{Burkert2020}, the fact that the observed number of GCs around DF2 is 10, is in excellent agreement with our tidal formation scenario. We emphasise, though, that this is fairly sensitive to the initial $R\sub{e,GC}$ adopted; if we set $R\sub{e,GC} = 3 \kpc$, rather than $2\kpc$, the average fraction of GCs that is stripped off increases to $\sim 80$ percent (Appendix~\ref{app:gc_orb}). 

We conclude that the tidal formation scenario for DMDGs like DF2 and DF4 is viable in that it can successfully explain the DM deficiency, the low central surface brightness, the low stellar velocity dispersion, as well as the phase-space properties of the GC population, without having to resort to an extreme orbit or extreme properties of the pre-infall progenitor galaxy. The only requirements are that the progenitor has a cored DM halo, which, as discussed above, is perfectly plausible, and that the initial distribution of GCs is relatively concentrated. A testable prediction of this tidal formation scenario is that, due to the tidal stirring-induced mixing, the outskirts of the tidal remnants should have very shallow metallicity gradients. The only puzzling observation of DF2 and DF4 that remains without proper explanation in this scenario is the abnormal luminosity functions of their GCs. Whereas it is to be expected that the tides preferentially strip the less massive GCs (the more massive ones experience stronger dynamical friction, causing them to sink towards the centre where they are better shielded from tidal stripping), which goes in the right direction to explain the overabundance of luminous GCs observed in DF2, this fails to explain why DF2 has multiple GCs that appear similar to $\omega$-Cen, the most luminous GC of the Milky Way. It remains to be seen whether a viable solution for this abnormality can be found within the tidal formation scenario proposed here.

\section*{Acknowledgements}

We thank the anonymous referee for providing insightful comments. GO acknowledges the Waterloo Centre for Astrophysics Fellowship for the support. FvdB is supported by the National Aeronautics and Space Administration through Grant No. 19-ATP19-0059 issued as part of the Astrophysics Theory Program. AB acknowledges support from the Excellence Cluster ORIGINS which is funded by the Deutsche Forschungsgemeinschaft (DFG, German Research Foundation) under Germany's Excellence Strategy-EXC-2094-390783311. Numerical simulations were performed on the Graham cluster operated by Compute Canada (\url{www.computecanada.ca}).


\section*{Data Availability}

The data underlying this article will be shared on reasonable request to the corresponding author.



\bibliographystyle{mnras}
\bibliography{df2_gc} 


\appendix

\section{Orbital parameters}
\label{app:orb_params}

The simulations presented in the main body of this paper adopt orbital parameters ($x\sub{c}=1.0$ and $\eta=0.3$) that were tuned such that the stellar body, at the end of the simulation (after $10\Gyr$), has an effective radius and line-of-sight velocity dispersion that are roughly in agreement with the measurements of DF2. We emphasise, though, that this choice of orbital parameters is not necessarily unique, and that there may well be other combinations of $x\sub{c}$ and $\eta$ that yield similar outcomes. Since the main point of this paper is to investigate whether tides can transform a normal galaxy into a DMDG that is somewhat similar to DF2, this potential degeneracy is of no concern.

In this appendix we briefly highlight how the results depend on the orbital circularity $\eta$, while keeping $x\sub{c}$, which controls the orbital energy, fixed at unity. Here we use the satellite for which the initial DM halo is cored ($m=1$), and we adopt the fiducial spatial distribution of GCs with $r\sub{gc}=2.5\kpc$.

\begin{figure}
    \begin{center}
        \includegraphics[width=0.4\textwidth]{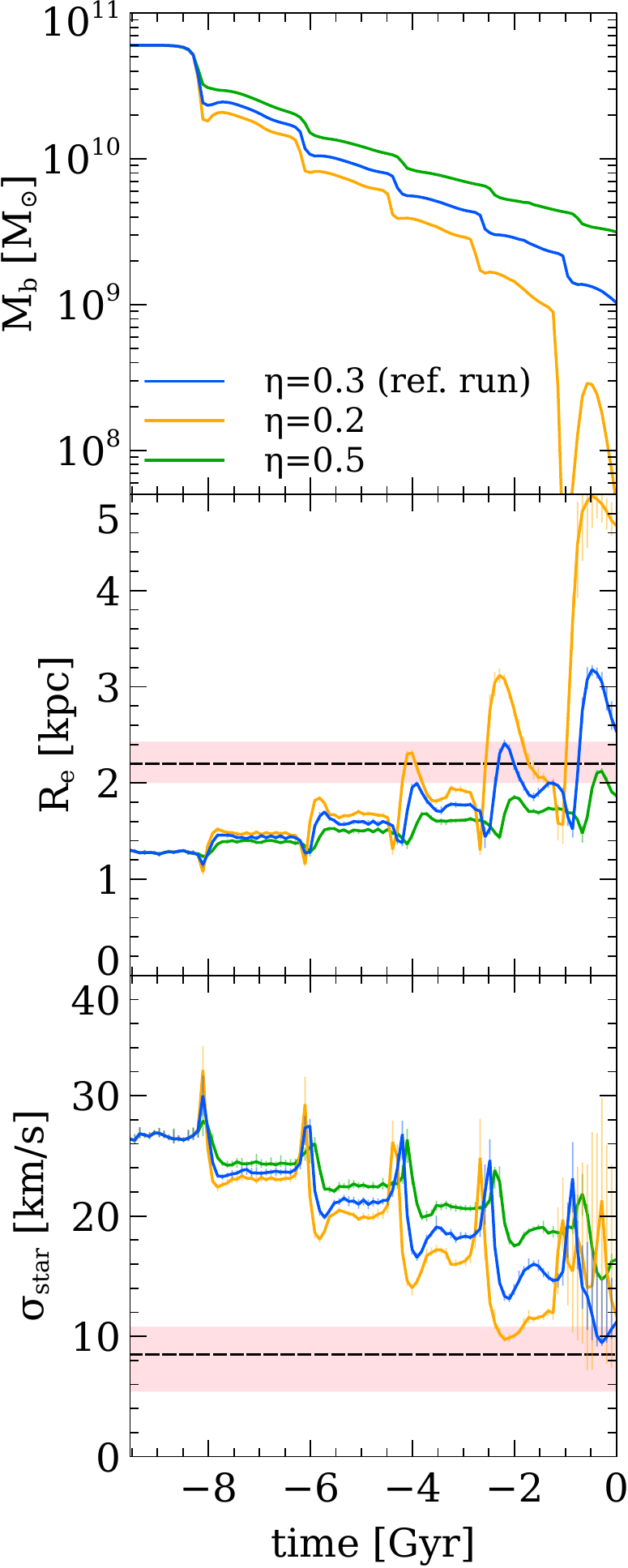}
    \end{center}
    \caption{Dynamical evolution of the \nbody satellite galaxy with a cored DM halo in {\tt Tide} simulations for different values of the orbital circularity, $\eta$, as indicated. All orbits have $x\sub{c}=1.0$. {\it Top panel:} evolution of the total bound mass (stars, DM and GCs) as a function of time. {\it Middle panel:} effective radius of the stellar body in projection, $R\sub{e}$, as a function of time. {\it Bottom panel:} line-of-sight velocity dispersion of stars, $\sigma\sub{star}$, as function of time. Horizontal dashed lines and shaded regions indicate the observed values for DF2 (see caption of \autoref{fig:galaxy} for details), and are shown for comparison. Note how more radial orbits (smaller $\eta$) leads to more mass loss and stronger tidal stirring (causing an increase in $R\sub{e}$ and a decrease in $\sigma\sub{star}$.
    \label{fig:interaction_orb}}
\end{figure}

\autoref{fig:interaction_orb} shows that when the satellite galaxy is on a more radial orbit, characterised by a smaller $\eta$, it experiences more significant tidal mass loss. This is simply a consequence of the fact that a smaller $\eta$ implies a smaller peri-centric distance, and thus stronger tides. Consequently, the satellite experiences more tidal stirring, causing it to `puff-up' more. The middle panel shows that in the case of $\eta = 0.2$, the effective radius at the end of the simulation is almost twice as large as in our reference case ($\eta=0.3$). In fact, as evident from the evolution of the bound mass (top panel) and the stellar velocity dispersion (bottom panel), the system is close to break-up. Using a more circular orbit, with $\eta = 0.5$, yields significantly less mass loss, a reduced increase in $R\sub{e}$, and a reduced decrease in $\sigma\sub{star}$. Consequently, at the end of the simulation it is still too compact and dynamically too hot to resemble DF2.

The results for the GC population (not shown) are similar to what is seen for the stars. In the simulation with $\eta = 0.2$, the GC orbits expand too much, and all GC end up being stripped from the satellite by the present time (i.e., after $10\Gyr$ of evolution). In the simulation with $\eta = 0.5$, tidal stirring is insufficient to produce a GC population with phase-space parameters that are consistent with DF2.
 
These results motivate the choice for $\eta=0.3$ adopted as the fiducial value throughout this paper.

\section{Initial extent of the population of globular clusters}
\label{app:gc_orb}

In this Appendix we briefly examine how the tidal evolution of the GC population depends on its initial, spatial extent. As discussed in \autoref{sssec:gc_satellite}, we initialise the GCs at a fixed radius, $r\sub{GC}$, from the centre of the satellite system, after which they quickly `disperse' into a spatial distribution that is well characterised by a \sersic profile with $n=0.5$ and an effective radius, $R\sub{e,GC}$, that is comparable to $r\sub{GC}$. Throughout the paper we adopt $r\sub{GC}=2.5\kpc$ as our fiducial value, which yields an initial $R\sub{e,GC}=2.0\kpc$. This is consistent with, but at the low end of the distribution of $R\sub{e,GC}$ values for our (initial) satellite galaxy. The average value of $R\sub{e,GC}$ is somewhat larger \citep[e.g.,][]{Forbes2017, Hudson2018}. In particular, \cite{Forbes2017} suggests that the average $R\sub{e,GC}$ for a system with a halo mass of $6\times 10^{10}\Msun$ is $3.1\kpc$. Through trial and error we find that if we initialise our GCs with $r\sub{GC}=4.0\kpc$, they quickly disperse to a $n=0.5$ \sersic profile with $R\sub{e,GC} \simeq 3.1\kpc$. 

\begin{figure}
    \begin{center}
        \includegraphics[width=0.4\textwidth]{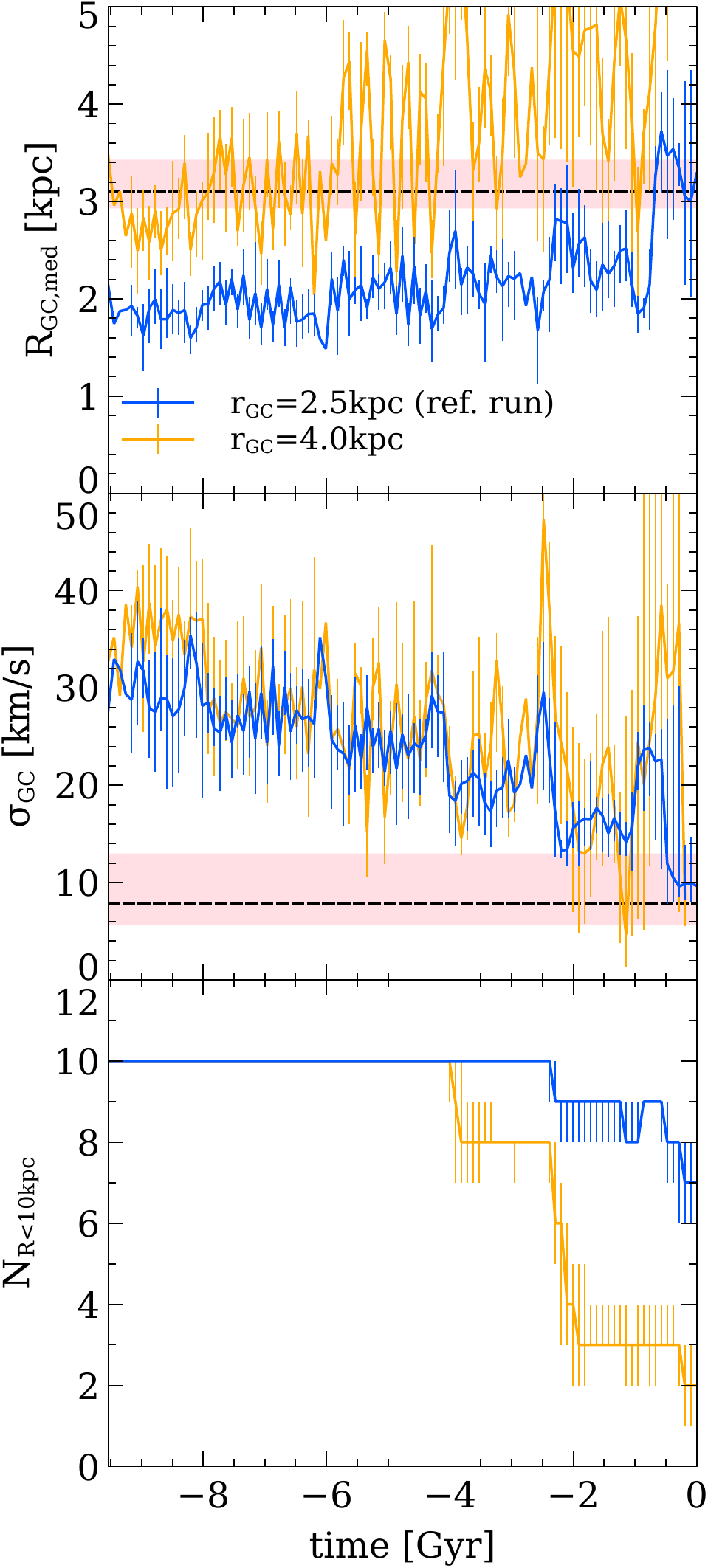}
    \end{center}
    \caption{Evolution of the GC population in {\tt Tide} simulations for different values of the initial radius of the GCs with respect to the centre of the satellite, $r\sub{GC}$, as indicated. All other parameters are kept fixed to those of our reference run. {\it Top panel:} the median, projected radius, $R\sub{GC,med}$ of the GCs with $R<10\kpc$ (a reasonable proxy for bound GCs). {\it Middle panel:} the line-of-sight velocity dispersion of the GCs with $R<10\kpc$. {\it Bottom panel:} the number of GCs with $R<10\kpc$. Errorbars indicate the 15-85 percentile range of values obtained using 100 random orientation angles. Horizontal dashed lines and shaded regions indicate the observed values for DF2 (see caption of \autoref{fig:gc} for details), and are shown for comparison. Note how the number of GCs that is stripped off from the satellite depends sensitively on the assumed value for $r\sub{GC}$.
    \label{fig:rgc}}
\end{figure}

\autoref{fig:rgc} compares the evolution of the GC population in two simulations that only differ in their value of $r\sub{GC}$. All other parameters are kept fixed to those of our reference run (i.e., $x\sub{c}=1.0$, $\eta=0.3$, and the initial DM halo is cored). Blue and orange lines show the results for $r\sub{GC}=2.5\kpc$ (our fiducial value) and $r\sub{GC}=4.0\kpc$, respectively. Whereas the evolution in $\sigma{GC}$ is similar in both cases (middle panel), the fraction of GCs that remain bound to the satellite (bottom panel) in the case with $r\sub{GC}=4.0\kpc$ is significantly reduced with respect to our reference run. In particular, on average only about 20 percent of the initial GCs remain bound (or, using our proxy for boundness, have $R<10\kpc$). Consequently, the initial satellite would have had to start out with $\sim 50$ GCs at infall in order to end up with 10 GCs (as in DF2) at the present time. Such a large number of GCs is unrealistic for a halo of $6 \times 10^{10}\Msun$ \citep[][]{Burkert2020}. Hence, in order to end up with a GC population similar to what is observed in DF2, the progenitor galaxy at infall needs to have a distribution of GCs that is fairly compact, but not unrealistically so.

Finally, we emphasise that the evolution of the effective radius and line-of-sight velocity dispersion of the stars is not significantly effected by a change in $r\sub{GC}$, indicating that the dynamical friction experienced by the GCs does not significantly heat the stellar component (i.e., the total stellar mass of the GCs is too small to impact the stellar body).

\bsp	
\label{lastpage}
\end{document}